%% file: main.tex
\documentclass[preprint,pteplogo]{ptephy_v2}

\preprintnumber{2505-18598} 
\usepackage{hyperref}
\usepackage{dirtytalk} 
\usepackage{comment}
\usepackage{bm}
\usepackage{nth}
\usepackage[flushleft]{threeparttable}
\usepackage{array,booktabs,makecell}
\usepackage[dvipsnames]{xcolor}
\usepackage{amsmath}

\begin{document}

\title{Measurement of $\Lambda$ Polarization \\in the $\PiKL$ Reaction at $p_{\pM}=1.33$ GeV/$c$ \\toward a New $\Lp$ Scattering Experiment}

\author[1]{T. Sakao \thanks{This work was carried out during Sakao's PhD course at the Department of Physics, Graduate School of Science, Tohoku University.} \thanks{Present address: Computer and Data Science Laboratories, NTT, Inc., Tokyo, Japan. Correspondence to: Tamao Sakao $<$tamao.sakao@ntt.com$>$}}
\author[1,3]{K. Miwa}
\affil{Department of Physics, Tohoku University, Sendai 980-8578, Japan}

\author{J. K. Ahn}
\affil{Department of Physics, Korea University, Seoul 02841, Korea}

\author{Y. Akazawa} 
\affil{Institute of Particle and Nuclear Studies, High Energy Accelerator Research Organization (KEK), Tsukuba 305-0801, Japan}

\author[1]{T. Aramaki}

\author{S. Ashikaga}
\affil{Department of Physics, Kyoto University, Kyoto 606-8502, Japan}

\author{S. Callier}
\affil{OMEGA Ecole Polytechnique-Centre National de la Recherche Scientifique IN2P3, 3 rue Michel-Ange, 75794 Paris 16, France}

\author[1]{N. Chiga}
\author[2]{S. W. Choi}

\author{H. Ekawa}
\affil{High Energy Nuclear Physics Laboratory, RIKEN, Wako 351-0198, Japan}

\author{P. Evtoukhovitch}
\affil{Joint Institute for Nuclear Research, Dubna, Moscow Region 141980, Russia}

\author[1]{N. Fujioka}

\author{M. Fujita}
\affil{Graduate School of Science, the University of Tokyo, 113-0033, Japan}

\author[4]{T. Gogami}
\author[4]{T. Harada}
\author{S. Hasegawa}
\affil{Advanced Science Research Center, Japan Atomic Energy Agency, Tokai, Ibaraki 319-1195, Japan}

\author[1]{S. H. Hayakawa}
\author[3]{R. Honda}

\author{S. Hoshino}
\affil{Department of Physics, Osaka University, Toyonaka 560-0043, Japan}

\author[9]{K. Hosomi}

\author[3,9]{M. Ichikawa}

\author[1]{Y. Ichikawa}
\author[3]{M. Ieiri}
\author[1]{M. Ikeda}
\author[9]{K. Imai}
\author[1]{Y. Ishikawa}
\author[3]{S. Ishimoto}
\author[2]{W. S. Jung}
\author[1]{S. Kajikawa}
\author[1]{H. Kanauchi}

\author{H. Kanda}
\affil{Research Center for Nuclear Physics, Osaka University, Ibaraki 567-0047, Japan}

\author[1]{T. Kitaoka}
\author[2]{B. M. Kang}

\author{H. Kawai}
\affil{Department of Physics, Chiba University, Chiba 263-8522, Japan}

\author{S. H. Kim}
\affil{Department of Physics, Kyungpook National University, Daegu 41566, Korea}
\author[10]{K. Kobayashi}
\author[1]{T. Koike}
\author[1]{K. Matsuda}
\author[1]{Y. Matsumoto}

\author[8]{S. Nagao}

\author[10]{R. Nagatomi}
\author[10]{Y. Nakada}
\author[6]{M. Nakagawa}
\author[3]{I. Nakamura}
\author[1]{T. Nanamura}
\author[4]{M. Naruki}
\author[1]{S. Ozawa}
\author[5]{L. Raux}
\author[1]{T. G. Rogers}
\author[10]{A. Sakaguchi}
\author[9]{H. Sako}
\author[9]{S. Sato}
\author[1]{T. Shiozaki}
\author[11]{K. Shirotori}
\author[4]{K. N. Suzuki}
\author[3]{S. Suzuki}
\author[12]{M. Tabata}
\author[5]{C. d. L. Taille}
\author[3]{H. Takahashi}
\author[3]{T. Takahashi}

\author{T. N. Takahashi}
\affil{Nishina Center for Accelerator-based Science, RIKEN, Wako, 351-0198, Japan}

\author[1,9]{H. Tamura}
\author[3]{M. Tanaka}
\author[9]{K. Tanida}
\author[15]{Z. Tsamalaidze}
\affil{Georgian Technical University, Tbilisi, Georgia}

\author[1,3]{M. Ukai}
\author[1]{H. Umetsu}
\author[1]{S. Wada}
\author[9]{T. O. Yamamoto}
\author[1]{J. Yoshida}
\author{K. Yoshimura}
\affil{Department of Physics, Okayama University, Okayama 700-8530, Japan}



\input{begin}
\input{abstract}
\input{introduction}
\input{experiment}
\input{analysis_flow}

\input{analysis1}

\input{analysis2}
\input{analysis3}

\input{analysis4}

\input{analysis5}

\input{systematic}

\input{results}
\input{discussion}

\clearpage
\input{summary}
\clearpage
\input{reference}

\input{end}

%% file: abstract.tex
\begin{abstract}%
Understanding the spectrum and dynamics of excited nucleon states ($N^{*}$ resonances) remains a central challenge in hadron physics, as these resonances emerge in intermediate states of $\pi N$ and $\gamma N$ interactions. The dynamical coupled-channel (DCC) models addressing this problem rely critically on experimental inputs for partial wave analyses (PWA). However, polarization observables of hyperons, which serve as sensitive probes of reaction mechanisms, have so far been limited in precision. 
To end this, we aim to provide high-precision measurements of the polarization of the $\Lambda$ hyperon produced in the $\PiKL$ reaction. The data were obtained with the J-PARC E40 experiment at the K1.8 beamline of the J-PARC Hadron Experimental Facility. The polarization of the $\Lambda$ was precisely measured in the angular range $0.6 < \costkz < 1.0$ with a fine bin width of $d\costkz = 0.05$. The observed average polarization in the region $0.60 < \costkz < 0.85$ was $0.932 \pm 0.058 (\text{stat}) \pm 0.028 (\text{syst})$. 
These results provide essential input for PWAs of DCC models, contributing to a deeper understanding of $N^{*}$ resonance dynamics. Furthermore, the strong $\Lambda$ polarization observed indicates the feasibility of developing a highly polarized $\Lambda$ beam, opening prospects for future $\Lambda p$ scattering experiments such as J-PARC E86.

\end{abstract}

\subjectindex{D14 Hypernuclei, D32 Hadron structure and interactions}

\maketitle

%% file: introduction.tex
\section{\bf Introduction}
The nucleon resonances, $N^*$, that appear as the intermediate states in $\pi N$ and $\gamma N$ reactions are direct manifestations of the non-perturbative dynamics of quarks and gluons (in the low-energy regime of QCD), providing crucial insights for understanding hadron structure and the strong interaction. The systematic investigation of the mass, width, and decay branching ratios of $N^*$ is termed \say{baryon spectroscopy.} By comparing these results with theoretical predictions from models such as the quark model and lattice QCD, it is possible to test the internal degrees of freedom within hadrons \cite{klempt-2010}. 

However, analysing a single channel (for example, only $\pi N\to \pi N$) presents the problem that the nature of the resonance cannot be correctly extracted \cite{svarc-2005} since the $N^*$ observed in experiments is often coupled to various channels, not only $\pi N$ but also $\eta N,\ K\Lam,\ K\Sigma,\ \pi\pi N$, and others. 

The dynamical coupled-channel (DCC) model is a theoretical framework that approaches this challenge by performing a \say{dynamically coupled} analysis of all major channels through a partial wave analysis (PWA) using experimental data, such as spin observables of the product particle \cite{matsuyama-2007, kamano-2010, Kamano-2016}. Regarding the $\PiKL$ reaction, the available experimental data, especially $\Lam$ polarization data \cite{Baker-1978, Saxon-1980}, have large uncertainties, resulting in the mechanism of the $N^*$ emerging not being fully understood yet. This paper reports the high-precision experimental data of $\Lam$ polarization in the $\PiKL$ reaction. Our data could be an essential input for the DCC models, which typically perform simultaneous fits to experimental data of spin observables in a wide range of reaction channels (e.g., $\pi N \to \eta N, K\Lam, K\Sigma$, and $\gamma N \to \pi N$) to extract the model parameters. 



This paper is also relevant to the study of hyperon-nucleon ($YN$) interactions, second only to nucleon-nucleon ($NN$) interactions.
The $NN$ interaction is classified as part of the baryon-baryon ($\BB$) interaction. Since this classification also includes interactions with hyperons, studying $\BB$ interactions leads to a more general understanding of the nuclear force as a baryon-baryon force. Furthermore, as the $\BB$ interaction is the basis for describing and understanding nuclear systems containing hyperons, such as hypernuclei and neutron stars, theoretical and experimental studies on the $\BB$ interaction have been intensively carried out. \cite{yoshimoto-2021, hayakawa-2021, ekawa-2019}. Unlike abundant $NN$ scattering data, $YN$ scattering data are still scarce owing to experimental difficulties due to the short lifetime of hyperons. Instead, the energy level structures of $\Lam$ hypernuclei have been used as probes to investigate the $\LN$ interaction \cite{Tamura-2005, ukai-2008, yamamoto-2015, esser-2015, gogami-2016, gogami-2016-2, yang-2018, gogami-2021}. Here, the $\Lam$ hyperon can penetrate the nuclear interior and form deeply bound hypernuclear states since it is not subject to Pauli blocking by other nucleons. Using hypernuclear data, two-body $\LN$ interaction theories have been updated \cite{hiyama-2010}. 

However, current theoretical models incorporating the current $\LN$ interaction fail to explain the existence of recently observed massive neutron stars \cite{Demorest-2010, Antoniadis-2013} (known as the \say{hyperon puzzle} of neutron stars). To establish more realistic $YN$ interaction models, it was necessary to obtain the precise two-body $YN$ interactions by scattering experiments. We (J-PARC E40 experiment \cite{Miwa-2019, Miwa-2020}) have obtained high-statistic $\Sp$ scattering data at the Japan Proton Accelerator Research Complex (J-PARC). Here, the differential cross-sections for $\SMp$ scattering \cite{Miwa-SMp}, $\SMpLn$ inelastic scattering \cite{Miwa-SMpLn}, and $\SPp$ scattering \cite{Nana-SPp} have been precisely measured. Ref. \cite{Nana-SPp} also derived the phase shifts for $\delta_{^{3}S_{1}}$ and $\delta_{^{1}P_{1}}$ of the $\SPp$ channel for the first time.

The modern description of the $YN$ interactions based on the chiral effective field theory ($\chi$EFT) constructed by \wip{J. Haidenbauer} \cite{Haidenbauer-2020} has recently updated their low-energy constants (LECs) with our data. The $YN$ interaction with strangeness $S=-1$ has advanced to the next-to-next-to-leading \wip{order} (NNLO) \cite{EFT-YN-NNLO}. However, the LECs have not yet been uniquely determined, and $\LN$ scattering data, particularly differential observables containing angular information such as differential cross-sections and spin observables, are indispensable. Therefore, we plan the next-generation $\Lp$ scattering experiment (J-PARC E86 experiment \cite{E86-proposal}) to collect $\Lp$ scattering data, including spin observables. As the spin observable, we will measure the analyzing power and depolarization of the $\Lam$ particle using a spin-polarized $\Lam$ beam produced by the $\PiKL$ reaction. Although the experimental design of J-PARC E86 is based on $\Lam$ polarization data in the $\PiKL$ reaction, measured in the 1980s \fix{\cite{Baker-1978, Saxon-1980}}, these data have rather large uncertainties, so this polarization should be reconfirmed with more precise data. Additionally, establishing a reliable method to measure $\Lam$ polarization is essential for the future spin observables analysis. To accurately verify the feasibility of J-PARC E86, we measured the $\Lam$ polarization using the $\PiKL$ reaction data collected during the $\SMp$ scattering data collection in J-PARC E40. 

The remainder of this paper is organized as follows: first, we describe the experimental setup of J-PARC E40 in Sec. \ref{sec-exp}, focusing on identifying $\Kz$ and $\Lam$ particles. After describing the analysis flow in Sec. \ref{sec-anaflow}, we explain the identification of the $\PiKL$ reaction and the $\Ldecay$ decay in Sec. \ref{sec-ana1} and Sec. \ref{sec-ana2}. The background subtraction and data correction are introduced in Sec. \ref{sec-ana3} and Sec. \ref{sec-ana4}. We then present the derivation of $\Lam$ polarization in Sec. \ref{sec-ana5} and the estimation of systematic uncertainty in Sec. \ref{sec-sysun}. Finally, we report the final result of $\Lam$ polarization in Sec. \ref{sec-result}. The discussion on $N^{*}$ resonances produced as intermediate states in $\PiKL$ reactions was described in Sec. \ref{sec-discussion}.

%% file: experiment.tex
\section{\bf Experiment} \label{sec-exp}
\begin{figure}[!h]
\centering\includegraphics[width=\columnwidth]{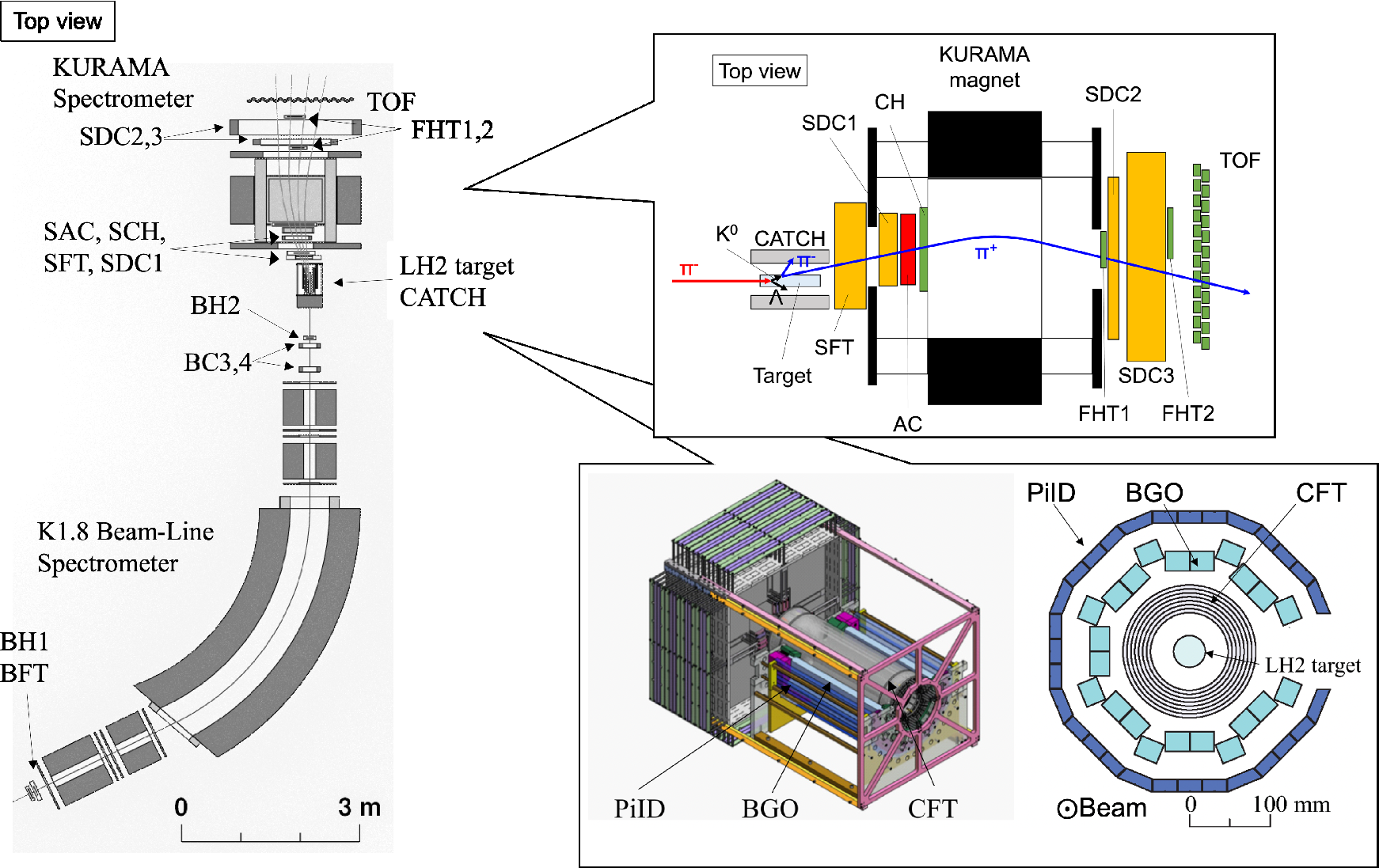}
\caption{Schematic view of the experimental setup in J-PARC E40. After the K1.8 beamline spectrometer, composed of QQDQQ magnets, a cylindrical detector cluster (CATCH \cite{Akazawa}) surrounding the liquid hydrogen (LH$_2$) target, and a forward magnetic spectrometer (KURAMA) were installed.}
\label{fig-e40setup}
\end{figure}

J-PARC E40 was conducted at the K1.8 beamline in the J-PARC Hadron Experimental Facility. Its data taking for $\SMp$ scattering was performed in February 2019, where a $\pM$ beam with a momentum of 1.33 GeV/$c$ was used for $\SM$ production. The beam intensity was $2.0\times10^{7}$ /spill, where the spill is a beam cycle of 5.2 seconds with a beam extraction of 2 seconds. The experimental setup is shown in Fig. \ref{fig-e40setup}.

The $\pM$ beam was transported to an experimental target through the K1.8 beamline, the most downstream of which is the K1.8 beamline spectrometer, composed of a QQDQQ magnet and position detectors, to measure the momentum of the beam particles. \wip{The QQDQQ magnetic system is designed to utilize point-to-point optics from the upstream reference (40 cm to the edge of the most upstream Q magnet) and downstream (30 cm from the pole edge of the most downstream Q magnet) points, so that the multiple scattering at the entrance and exit of the QQDQQ does not affect the momentum resolution to the first order. The momentum resolution is expected to be $3.3\times10^{-4}$ (FWHM) \cite{Takahashi}.} 

The cylindrical detector cluster (CATCH) surrounding the liquid hydrogen target (LH$_2$) and a forward magnetic spectrometer (KURAMA) were placed downstream of the K1.8 beamline spectrometer. CATCH \cite{Akazawa}, consisting of the LH$_2$ target, a cylindrical fiber tracker (CFT), a BGO calorimeter, and a scintillation fiber hodoscope (PiID), measures the kinetic energy of the scattered particles. \wip{The energy calibrations of CFT and BGO were performed using the $pp$ elastic scattering data with the momentum range of 0.45-0.85 GeV/$c$, referring to the scattering angle and energy deposit of protons \cite{Miwa-SMp}. The efficiencies of CFT and BGO were also estimated using these $pp$ scattering data.}

The KURAMA spectrometer consists of a dipole magnet (KURAMA magnet), tracking detectors (SFT, SDC1, SDC2, SDC3, FHT1, and FHT2), and trigger counters (SAC, SCH, and TOF). Trajectories of scattered particles at the entrance and exit of the KURAMA magnet were measured using these tracking detectors. The flight path of scattered particles in the magnetic field was determined through numerical iteration using the Runge-Kutta method \cite{Runge}. The estimated momentum resolution of the KURAMA spectrometer is $\Delta p/p\sim10^{-2}$ \cite{Takahashi} with a magnetic field of 0.78 T at the center. The time-of-flight (TOF) of the outgoing particle, along the flight path of $\sim3$ m, was also measured.  

\wip{An aerogel Cherenkov counter (SAC) with a refractive index of 1.10 was installed behind SDC1 to reject scattered $\pP$s at the trigger level}. Its efficiency was about 99\%, which allowed $\sim$1\% of all $\pi$s scattered to the KURAMA spectrometer mixed in its hit pattern. In addition, we also collected data requiring hits of SAC and TOF with a factor of 1/5000. These factors led to the $\PiKL$ reaction, in which $\Kz$ decays into $\pP\pM$, accumulating during $\SMp$ scattering data acquisition.

%% file: analysis_flow.tex
\section{\bf Analysis Flow}
\label{sec-anaflow}

The $\Lam$ polarization (hereafter referred to as $P_{\Lam}$) in the $\PiKL$ reaction can be measured \fixxx{using} the emission angle distribution of the decay proton from the $\Ldecay$ decay at the $\Lam$ rest frame \fix{for the $\PiKL$ reaction plane}. The \fix{angular} distribution \fix{of the decay proton} is defined as 
 \begin{equation}
  \frac{dN}{d\costp} = \frac{N_0}{2}(1+\alpha P_{\Lam} \costp), 
  \label{eq-costp}
\end{equation}
where $\costp$ is the emission angle of the proton, $N$ is the counts in each bin ($d\costp$), $N_0$ is the total counts \fix{integrating over all} $\costp$, and $\alpha$ is the asymmetry parameter of $\Ldecay$ decay ($\alpha = 0.750\pm0.009\pm0.004$ \cite{Alpha}). \fixxx{When} \fix{the} $\costp$ \fix{distribution} was measured, \fixxx{we could} \fix{obtain the value of $P_{\Lam}$ by fitting} it with the function corresponding to Equation (\ref{eq-costp}). The analysis procedures we used are briefly summarized as follows.

First, $\Kz$ in the $\PiKL$ reaction was reconstructed by detecting $\pP$ and $\pM$ from $\kzdecay$ decay. \fixxx{Subsequently}, $\Lam$ was identified \fixxx{using} the missing mass method \fix{using $\pM$ beam momentum information}. In the $\Kz$ reconstruction, $\pP$ was detected \fixxx{using} KURAMA \fixxx{with reference} to the correlation between mass \fix{sqaured} and momentum (\fix{Fig.} \ref{fig-pKurama_m2}), and $\pM$ was detected \fixxx{using} CATCH \fixxx{with reference} to the correlation between energy deposit measured \fixxx{using} CFT ($\Delta E$) and total energy measured \wip{using} BGO (\fix{Fig.} \ref{fig-dE_E}). Details \fixxx{are given} in Sec. \ref{sec-ana1}.

Second, the $\Ldecay$ decay was identified by detecting the decay proton and $\pM$ \fixxx{using} CATCH. Here, we checked the kinematic consistency for the magnitude and the \fix{direction} of \fix{the} $\Lam$ momentum vector, assuming the production ($\PiKL$) and decay ($\Ldecay$) kinematics. Finally, we identified $3.56\times10^{4}$ $\Lam$s (\fix{Fig.} \ref{fig-mm_aftercut}). The remaining background was \fixxx{primarily} caused by the multiple $\pi$ production ($\multipi$). Details \fixxx{are given} in Sec. \ref{sec-ana2}.

Third, \fix{we obtained the $\costp$ distribution and estimated background contribution contained therein}. Here, we referred to the $\Kz$ \fix{flight} length to separate the $\Lam$ production and background. The \fix{flight} length should be distributed following its lifetime. In contrast, \fix{the flight} length should distribute around zero if it was mistakenly calculated \fixxx{using} $\pP$ and $\pM$ from the $\multipi$ reaction. We subtracted the estimated background from $\costp$ of events within the $\Lam$ peak range in the missing mass spectrum. Details \fixxx{are available} in Sec. \ref{sec-ana3}.

Fourth, we corrected the extracted $\costp$ \fixxx{using} CATCH acceptance estimated \fixxx{using} a Geant4-based Monte Carlo simulation. \fixxx{Subsequently}, the \fix{original} $\costp$ \fix{distribution for the} $\Lam$ \fix{decay} was obtained in the \fix{production angular range of} $0.6<\costkz<1.0$ with an angular bin width of $d\costkz=0.05$. Details \fixxx{are given} in Sec. \ref{sec-ana4}.

Fifth, we optimized the angular range, which is to be used for the fitting, \fixxx{through} a chi-square test. Details \fixxx{are given} in Sec. \ref{sec-ana5}.

Finally, we fitted $\costp$ with a linear function corresponding to Equation (\ref{eq-costp}) to derive $P_{\Lam}$. The systematic uncertainties are explained in Sec. \ref{sec-sysun}. 

\wip{The present results and discussion are presented in Sec. \ref{sec-result} and Sec. \ref{sec-discussion}.}

%% file: analysis1.tex
\section{\bf Analysis \rom{1}: Identification of the $\PiKL$ reaction}
\label{sec-ana1}
The $\Lam$ polarization ($P_{\Lambda}$) analysis comprised three parts. \wip{(1) the $\Lam$ production identification by analyzing the missing mass of $\PiKX$ reaction, (2) the $\Ldecay$ decay identification, and (3) the $P_{\Lambda}$ derivation.} \wip{Hereafter, the missing mass of $\PiKX$ reaction will be referred to as $\mm$.}  This section describes the first component; the $\Lam$ production identification.

The momentum of the $\pM$ beam was reconstructed event by event using the K1.8 \wip{beamline} spectrometer. Details of this analysis are \wip{described} in Ref. \cite{Tamao-JPS}. The outgoing $\Kz$ was reconstructed by detecting the $\pP$ and $\pM$ with the KURAMA and CATCH, respectively. 

\begin{figure}[!h]
\centering\includegraphics[width=0.7\columnwidth]{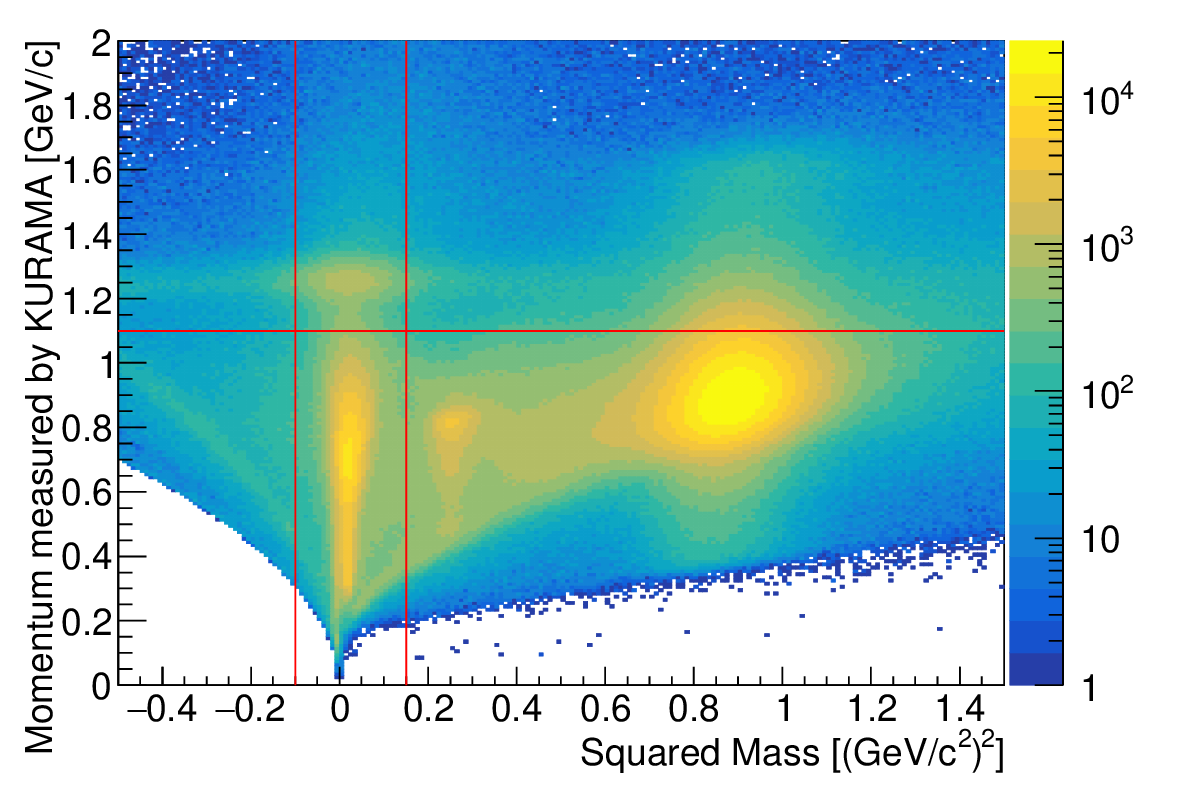}
\caption{Correlation between the mass squared ($m^{2}$) and momentum measured \fixxx{using} the KURAMA spectrometer, taken in E40 data. The region with $-0.1<m^{2}<0.15$ and $p<1.1$ was selected as $\pP$.}
\label{fig-pKurama_m2}
\end{figure}

The $\pP$ was identified using momentum and \fixxx{TOF} information. \fix{Fig.} \ref{fig-pKurama_m2} shows the correlation between the mass squared and momentum, measured \fixxx{using} the KURAMA spectrometer. The red solid lines represent the selected $\pP$ region.  

\begin{figure}[!h]
\centering\includegraphics[width=0.7\columnwidth]{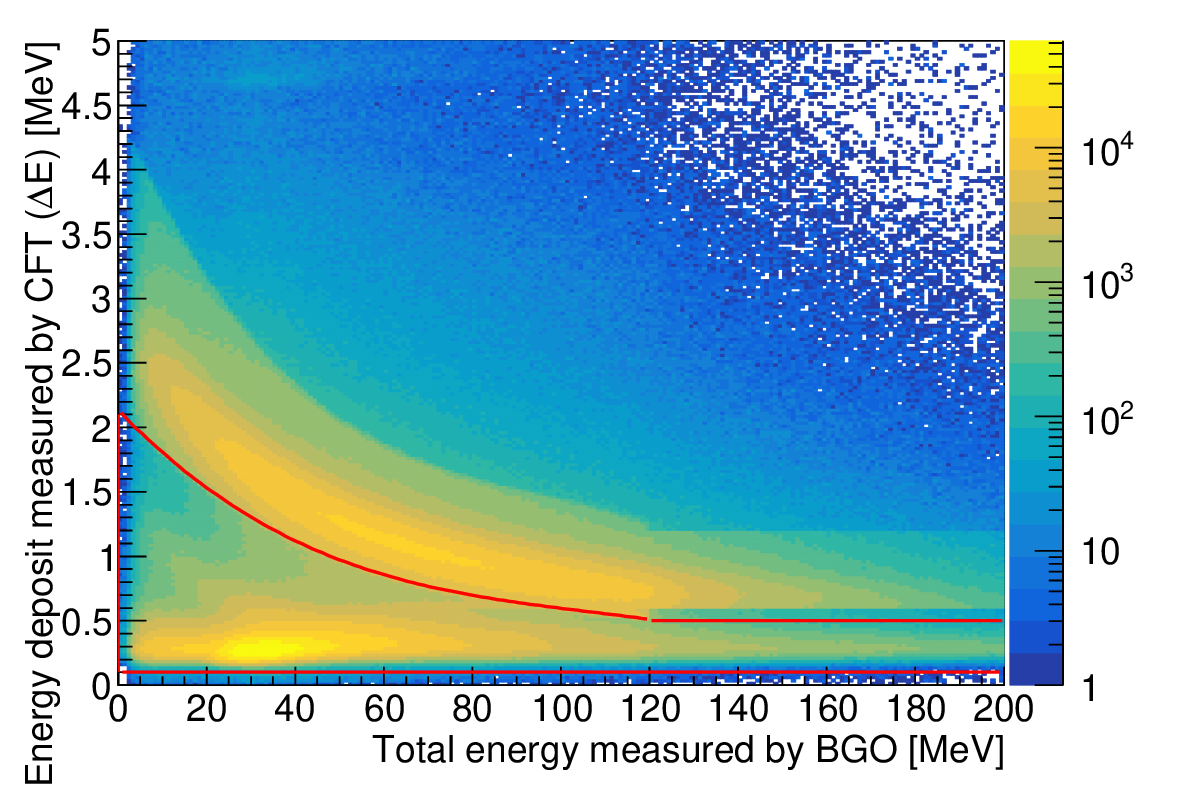}
\caption{Correlation between energy deposit measured \fixxx{using} CFT ($\Delta E$) and total energy measured \fixxx{using} BGO ($E$), taken in E40 data. The region surrounded by the red solid lines was selected as $\pM$.}
\label{fig-dE_E}
\end{figure}

The $\pM$ was identified using the so-called $\Delta E$-$E$ correlation between the energy deposition measured in CFT ($\Delta E$) and the total energy measured in BGO ($E$). Fig. \ref{fig-dE_E} shows the $\Delta E$-$E$ correlation after the CATCH calibration, highlighting the loci corresponding to protons and $\pi$s. \wip{Hereafter, the $\pi$s detected in CATCH were assumed as $\pM$s considering the charge conservation.} The kinetic energy of $\pM$ cannot be measured in BGO because many of $\pi$s \wip{pass through} it. Therefore, the momentum magnitude of $\pM$ was determined so that the invariant mass of $\pP$ and $\pM$ becomes the $\Kz$ mass.

The $\Kz$ momentum vector was reconstructed \fixxx{using} \fix{momentum} vectors of $\pP$ and $\pM$ ($\bm{P_{\pP}},\ \bm{P_{\pM}}$) assuming the $\kzdecay$ decay as
\begin{equation}
  \bm{P_{\Kz}} = \bm{P_{\pP}} + \bm{P_{\pM}}.
\end{equation}

\begin{figure}[!h]
\centering\includegraphics[width=0.7\columnwidth]{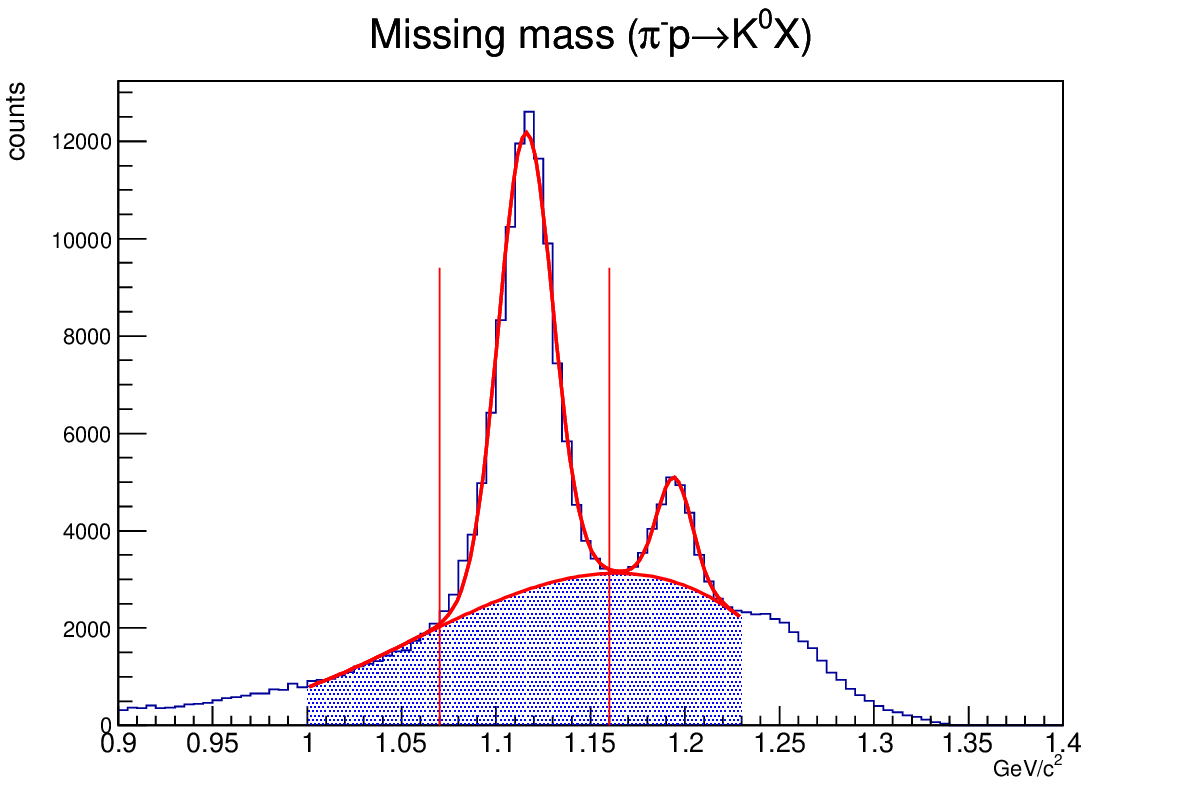}
\caption{Missing mass of the $\PiKX$ reaction, $\mm$. The red solid line represents the fitting function of two Gaussians and a third-order polynomial, and the blue shade represents the estimated background structure. The $\pm3\sigma$ interval of the first Gaussian (1.07-1.16 GeV/$c^{2}$) was defined as the $\Lam$ range for later analyses. The total yield of $\Lam$ was estimated to be $6.73\times10^{4}$.}
\label{fig-mm}
\end{figure}

Finally, $\Lambda$ was identified \fixxx{using} the missing mass method of the $\PiKX$ reaction, \wip{requiring the detection of proton and $\pM$ from the $\Ldecay$ decay.} Fig. \ref{fig-mm} shows the missing mass, $\mm$, with a fitting function of two Gaussians and a third-order polynomial. The left peak corresponds to $\Lam$, and the another $\Sigma^{0}$. The background structure (blue shade) originated from multiple $\pi$ production (i.e., $\multipi$). The $\pm3\sigma$ interval of the first Gaussian (1.07-1.16 GeV/$c^{2}$) was defined as the $\Lam$ range for later analyses. Here, $6.73\times10^{4}$ $\Lam$ particles were accumulated. 


%% file: analysis2.tex
\section{\bf Analysis \rom{2}: Identification of the $\Ldecay$ decay}
\label{sec-AnaLdecay}
\label{sec-ana2}
\wip{This section explains the second component of the $P_{\Lam}$ analysis, $\Ldecay$ decay identification. To correctly determine the cut conditions for the $\Ldecay$ decay identification, we simulated the $\PiKL$ reactions, normalized to the actual yield, followed by the $\Ldecay$ decays without the main background (i.e., multiple $\pi$ production). This simulation satisfied the following features; (1) all components of CATCH and KURAMA spectrometers were installed, (2) the measured energy resolution was applied to CFT and BGO, (3) the electromagnetic and hadronic processes were included as physics processes, and (4) The $\PiKL$ reaction was generated according to the angular distribution of previous experimental data \cite{Baker-1978}. Again, it is reasonable that only the $\Lam$ production reaction (i.e., the $\PiKL$ reaction followed by the $\Ldecay$ decay) was simulated, as the aim of this simulation is only to study the cut condition for $\Lam$ event selection.} 

The kinetic energy of the decay $\pM$ cannot be measured because the BGO thickness is insufficient, as mentioned in the explanation of $\Kz$ momentum reconstruction, whereas that of the decay proton can be. Therefore, the $\pM$ momentum magnitude was determined such that the invariant mass of the proton and $\pM$ becomes $\Lam$ mass using the momentum vector of the decay proton and the direction of $\pM$ momentum. Subsequently, the $\Lam$ momentum was obtained by summing the momentum vectors of the decay proton and $\pM$. 
 
The $\Ldecay$ decays were identified by checking the kinematic consistency for magnitude and \fix{direction} of the $\Lam$ momentum vectors \fix{constructed from two independent analyses, that is,} the production ($\PiKL$) and decay ($\Ldecay$) analyses. The \fix{angular difference ($\dtheta$)} of the \fix{same} momentum vectors obtained \fixxx{using} these different analyses was used as the valuation index. \fix{$\dtheta$ is calculated as}
\begin{equation}
    \fix{\cos{\dtheta}} = \frac{\bm{p_{production}} \cdot \bm{p_{decay}}}{|\bm{p_{production}}| \cdot |\bm{p_{decay}|}},
    \label{eq-costcm_1p1pi}
\end{equation}
where $\bm{p_{production}}$ is the $\Lam$ momentum vector \fix{obtained as the missing momentum of the $\PiKX$ reaction}, and $\bm{p_{decay}}$ is the $\Lam$ momentum vector obtained \fixxx{using} the above CATCH analysis. $\cos{\dtheta}$ should ideally be \fixxx{1}, as we here calculate the same momentum ($\Lam$) \fixxx{using} two different analyses. The difference between $\bm{p_{production}}$ and $\bm{p_{decay}}$, $\Delta p$, was used as another valuation index and calculated as
\begin{equation}
  \Delta p = |\bm{p_{decay}}| - |\bm{p_{production}}|.
\end{equation}
$\Delta p$ should ideally be 0. 

\begin{figure}[!h]
    \centering
    \includegraphics[width=0.7\columnwidth]{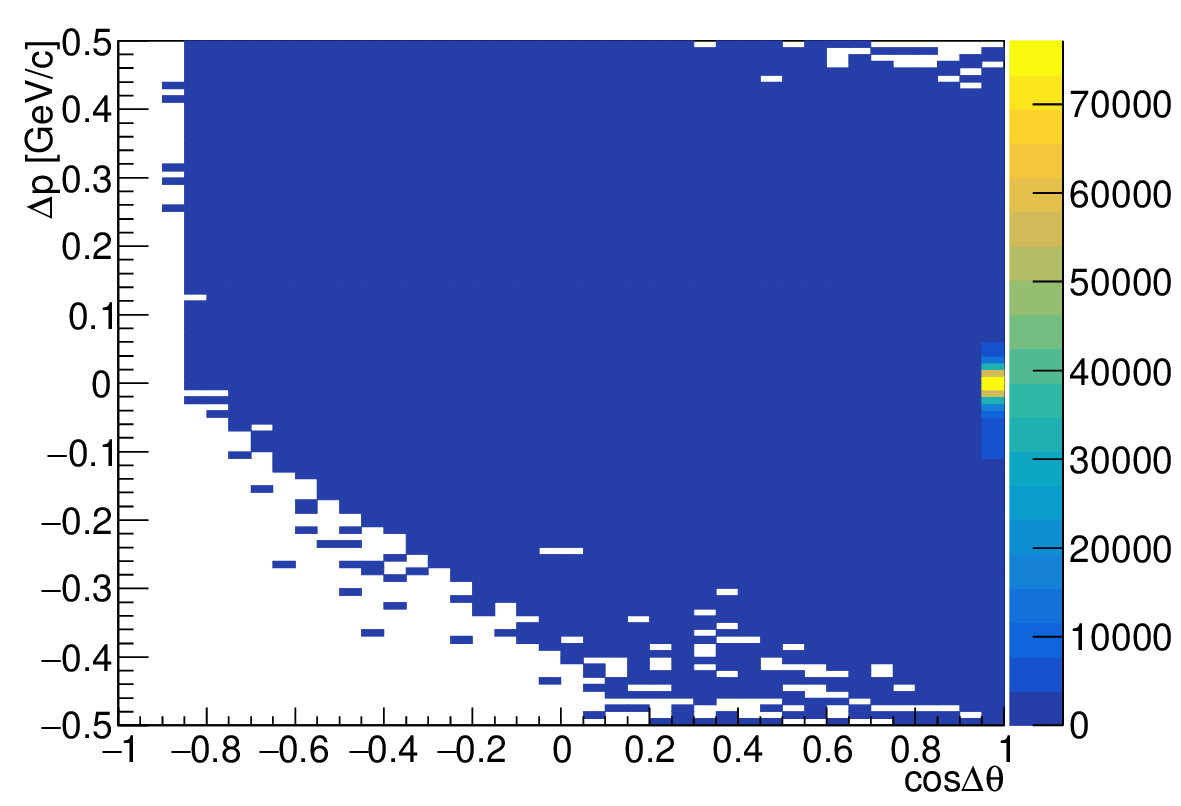} 
    \caption{Correlation between $\Delta p$ and $\cos{\dtheta}$, taken in the simulation data.}
    \label{fig-dpcost-sim}
\end{figure}

\begin{figure}[!h]
    \centering
    \includegraphics[width=0.7\columnwidth]{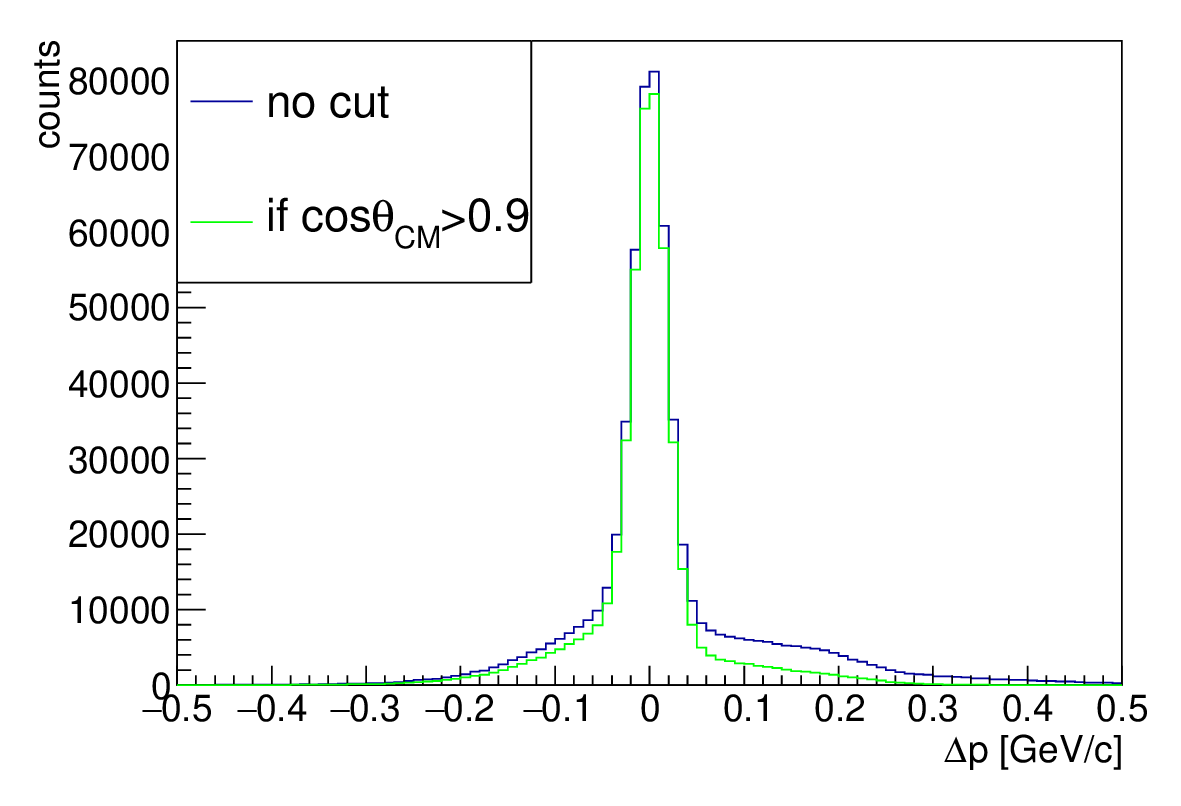} 
    \caption{$\Delta p$, taken in the simulation data. The green spectrum represents the $\Delta p$ after applying the $\cos{\dtheta}$ cut.}
    \label{fig-dp-sim}
\end{figure}

\begin{figure}[!h]
    \centering
    \includegraphics[width=0.7\columnwidth]{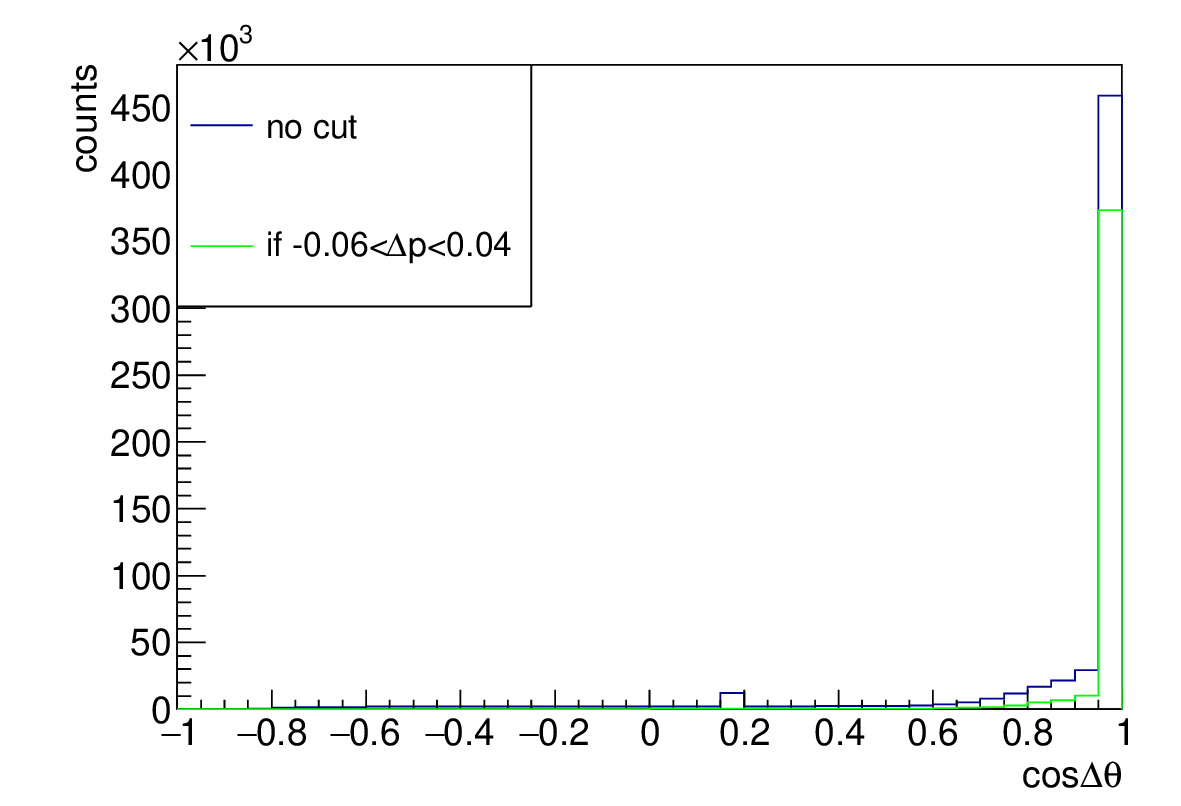} 
    \caption{$\cos{\dtheta}$, taken in the simulation data. The green spectrum represents the $\cos{\dtheta}$ after applying the $\Delta p$ cut.}
    \label{fig-cost-sim}
\end{figure}

\begin{figure}[!h]
    \centering
    \includegraphics[width=0.7\columnwidth]{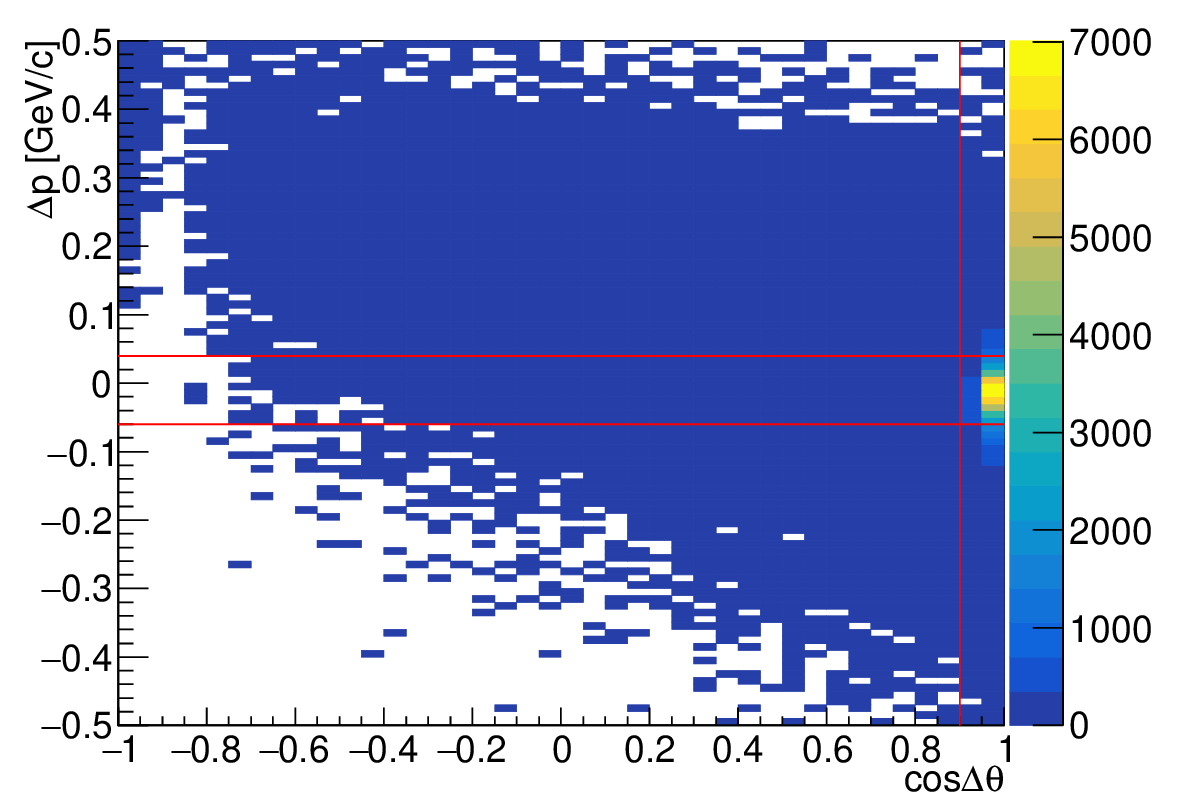} 
    \caption{Correlation between $\Delta p$ and $\cos{\dtheta}$, taken in \fixxx{the} E40 data. The red solid lines represent the cuts to select the $\Ldecay$ decay.}
    \label{fig-dpcost}
\end{figure}

\begin{figure}[!h]
    \centering
    \includegraphics[width=0.7\columnwidth]{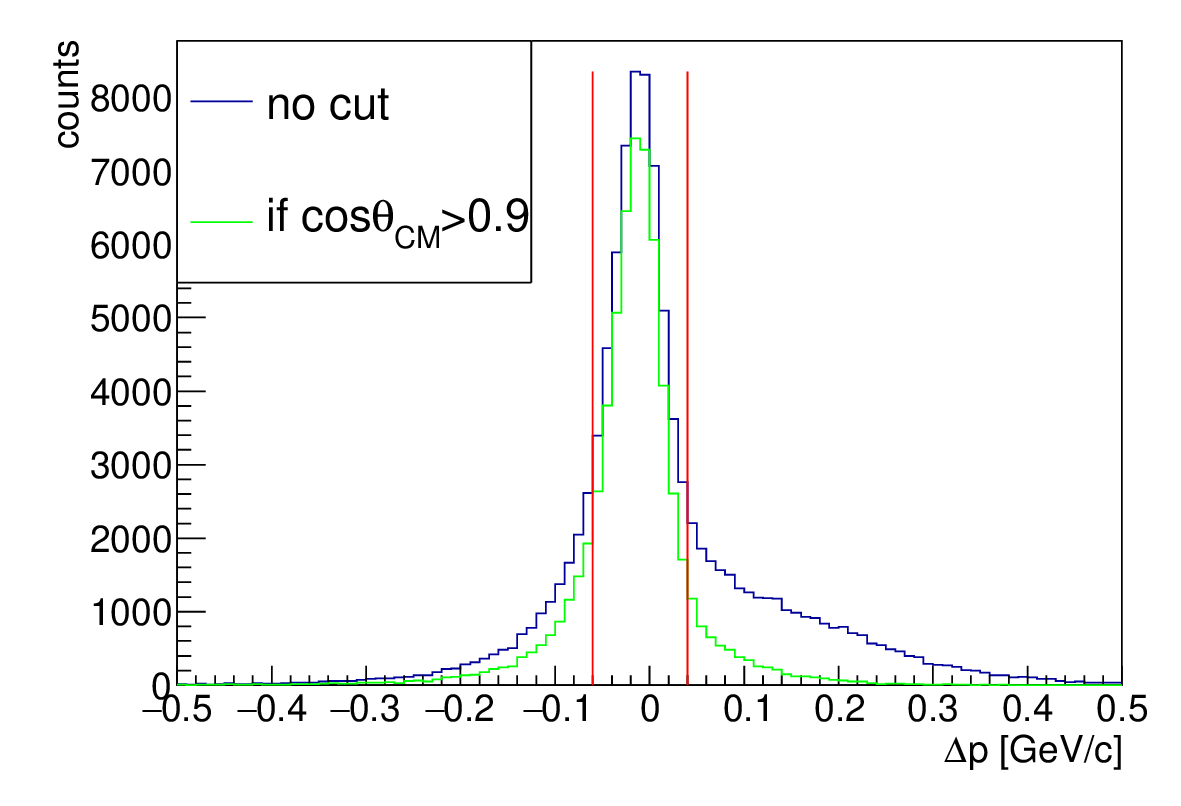} 
    \caption{$\Delta p$, taken in \fixxx{the} E40 data. The red solid lines represent the cuts to select the $\Ldecay$ decay. The green spectrum represents the $\Delta p$ after applying the $\cos{\dtheta}$ cut.}
    \label{fig-dp}
\end{figure}

\begin{figure}[!h]
    \centering
    \includegraphics[width=0.7\columnwidth]{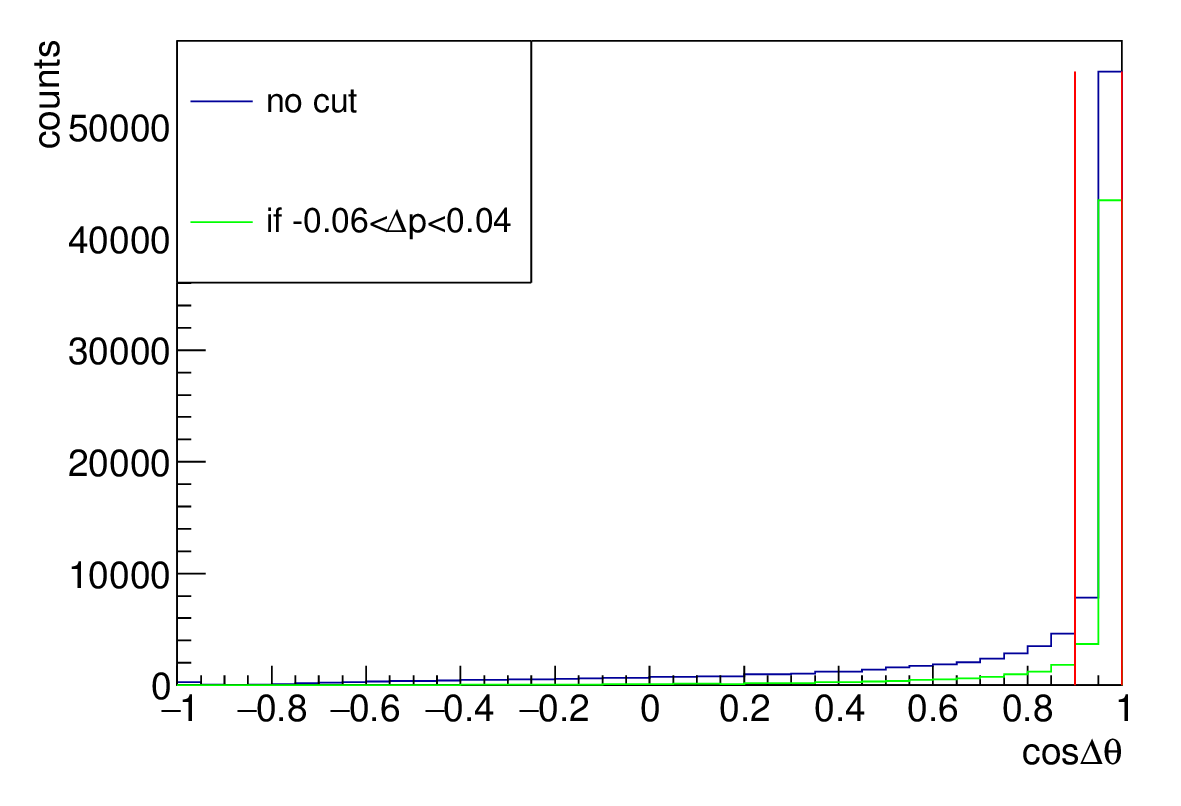} 
    \caption{$\cos{\dtheta}$, taken in \fixxx{the} E40 data. The red solid lines represent the cuts to select the $\Ldecay$ decay. The green spectrum represents the $\cos{\dtheta}$ after applying the $\Delta p$ cut.}
    \label{fig-cost}
\end{figure}

Figs. \ref{fig-dpcost-sim}, \ref{fig-dp-sim}, and \ref{fig-cost-sim} show the simulation results of the correlation between $\cos{\dtheta}$ and $\Delta p$ for $\Lam$ events, its $y$-projection (blue solid line), and its $x$-projection (top right, blue solid line), respectively. 
\wip{The reason for the broad distribution of Fig. \ref{fig-dp-sim} is assumed to be that hadron reactions occurring inside the BGO made the energy measurement incorrect, or that not only protons but also $\pM$s are detected in the same BGO segment, resulting in the energy measurement being wrong.} 
Fig. \ref{fig-dpcost}, \ref{fig-dp}, and \ref{fig-cost} show the same plots for E40 data. \wip{The peak corresponding to the $\Ldecay$ decay can be identified, in agreement with the simulation. However, due to the incomplete calibration of the CATCH and KURAMA spectrometer, the center of $\Delta p$ (Fig. \ref{fig-dp}) was off from the zero origin. Therefore, the cut region for $\Delta p$ was chosen as the region of $\pm0.05$ GeV/$c$ from the $\Delta p$ peak at $-0.01$ GeV/$c$.} 
Based on the simulation results of $\Delta p$ and $\cos{\Delta\theta}$, cut conditions were applied at $\cos{\dtheta}>0.9$ and $-0.06<\Delta p<0.04$ GeV/$c$, as represented by the red solid lines to select $\Lam$ events and suppress the multiple $\pi$ background. \sfix{The $\cos{\dtheta}$ and $\Delta p$ spectra after applying these cuts are represented by the green solid lines in Figs. \ref{fig-dp-sim}, \ref{fig-cost-sim}, \ref{fig-dp}, \fixxx{and} \ref{fig-cost}}. 

\begin{figure}[!h]
    \centering
    \includegraphics[width=0.7\columnwidth]{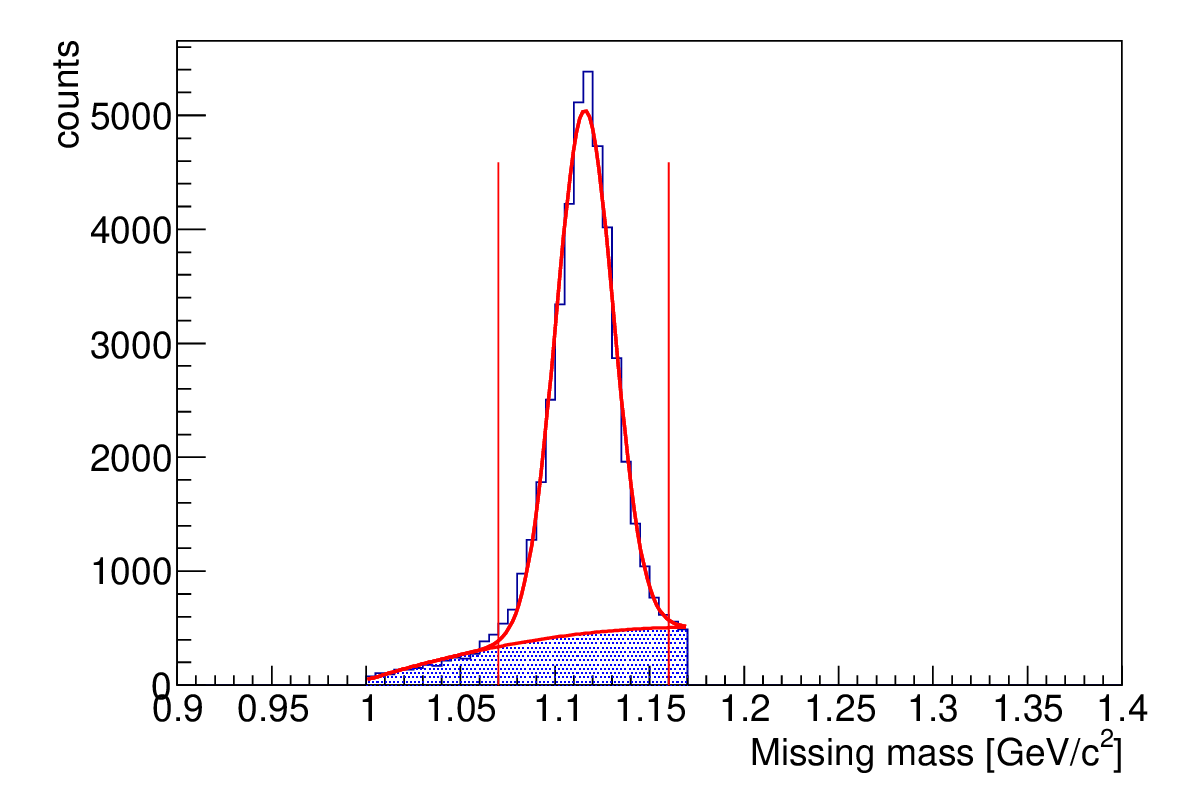}
    \caption{Missing mass \fix{spectrum} of the $\PiKX$ reaction, $\mm$, after \fix{applying the cut to select} the $\Ldecay$ decay. By fitting this with a Gaussian and a \fixxx{second} polynomial, we identified $3.56\times10^{4}$ $\Lam$s with an S/N of 4.37.}
    \label{fig-mm_aftercut}
\end{figure}

$\mm$ spectrum after selecting the $\Ldecay$ decay is shown in \fix{Fig.} \ref{fig-mm_aftercut}. The background from multiple $\pi$ production was suppressed well. By fitting this with a Gaussian and a \nth{2} polynomial, we identified $3.56\times10^{4}$ $\Lam$s ($S/N = 4.37$) in the $\pm3\sigma$ region (red solid lines). 


%% file: analysis3.tex
\section{\bf Analysis \rom{3}: Background Subtraction in the $\costp$ Distribution}
\label{sec-bgsubt}
\label{sec-ana3}

\begin{figure}[!h]
    \centering
    \includegraphics[width=0.7\columnwidth]{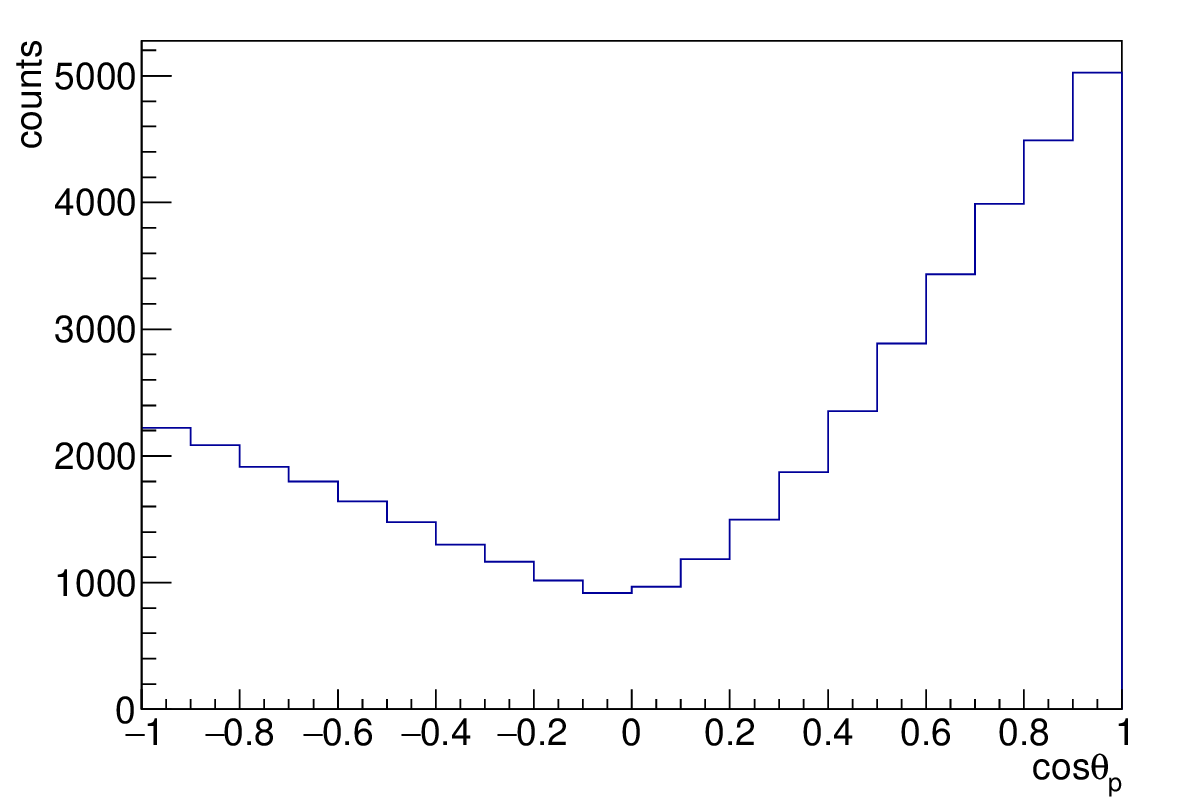}
    \caption{The $\costp$ distribution for the events within the range of $1.07<\mm<1.16$ GeV/$c^{2}$ (indicated by the red solid lines in Fig. \ref{fig-mm_aftercut}).}
    \label{fig-costp_all}
\end{figure}

\fix{To measure $P_{\Lam}$ using Equation (\ref{eq-costp}), we obtained the $\costp$ distribution for the events within the range of $1.07<\mm<1.16$ GeV/$c^{2}$ (indicated by the red solid lines in Fig. \ref{fig-mm_aftercut}), as shown in Fig. \ref{fig-costp_all}. \fixxx{We} can easily \fixxx{observe} a clear asymmetry in this $\costp$ distribution attributed to the decay mechanism of $\Lam$. The acceptance of CATCH led to a decrease in $\costp=0$. The problem is that this $\costp$ distribution still contains unwanted angular distribution due to background contamination in the range of $1.07<\mm<1.16$ GeV/$c^{2}$. We should remove this background contribution from the $\costp$ distribution of Fig. \ref{fig-costp_all}.} 
\fix{The background is the $\pP\pM$ pair, from multiple $\pi$ production ($\multipi$). To subtract the background contribution, we use the difference of the reaction vertex in the beam direction between multiple $\pi$ production and $\Kz$ production. Here, we define \fixxx{the} $\Kz$ flight length ($\dzkz$) as the distance between the production ($z_{production}$) and decay ($z_{decay}$) points of $\Kz$ with the following expression: }
\begin{equation}
  \dzkz = z_{decay} - z_{production}.
\end{equation}

For \fixxx{a} true $\Kz$ event, the $\dzkz$ distribution is asymmetric with an exponential tail toward larger values, corresponding to the $\Kz$ lifetime (\fixxx{the} $\Kz_s$ lifetime is $8.95\times10^{-11}$ \fixxx{s; thus}, $c\tau\sim2.69$ cm). In contrast, for the background events, the same distribution is symmetric with a peak around \fixxx{0}, as the production and decay points should be identical for events induced by the strong interaction.

\begin{figure}[!h]
    \centering
    \includegraphics[width=	0.8\columnwidth]{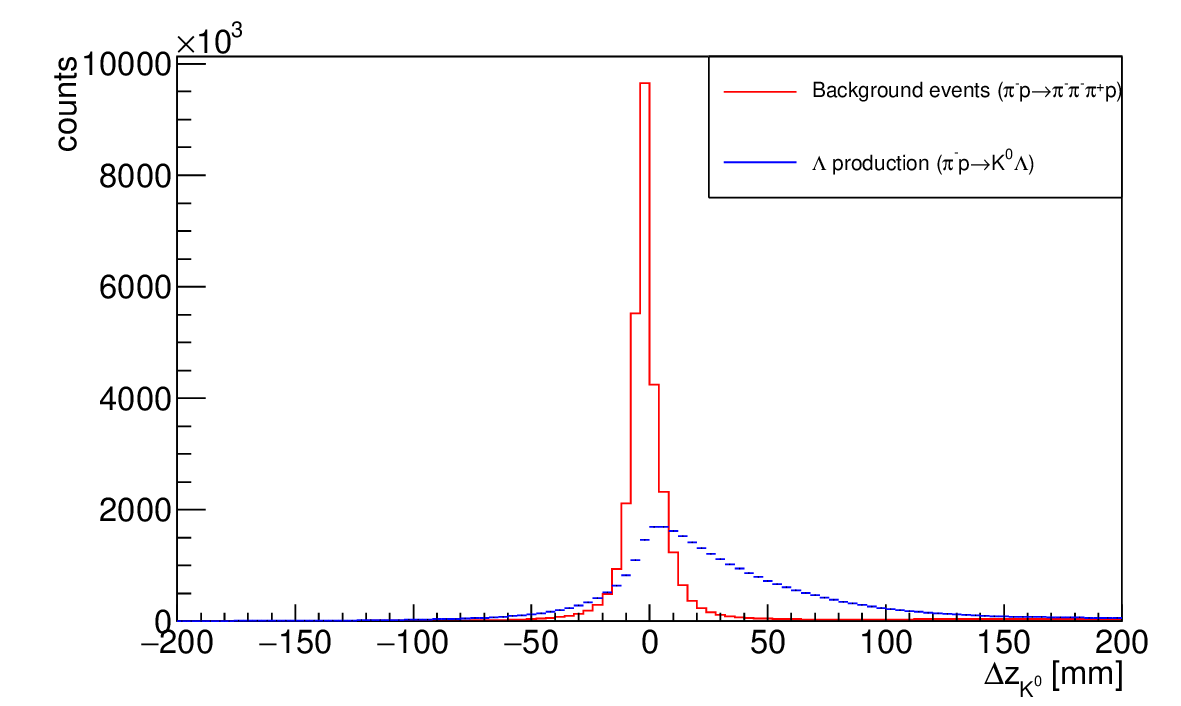}
    \caption{$\Kz$ \fix{flight} length in the beam direction ($\dzkz$) obtained in the simulations \fix{for the} background events (red) and \fix{the} $\Lam$ production (blue). The total counts are normalized.}
    \label{fig-dz_sim}
\end{figure}

\fix{Fig.} \ref{fig-dz_sim} shows $\dzkz$ obtained \fix{from} the simulation \fix{for the} background events (red) and $\Lam$ production (blue) with normalized total counts.  

\begin{figure}[!h]
    \centering
    \includegraphics[width=\columnwidth]{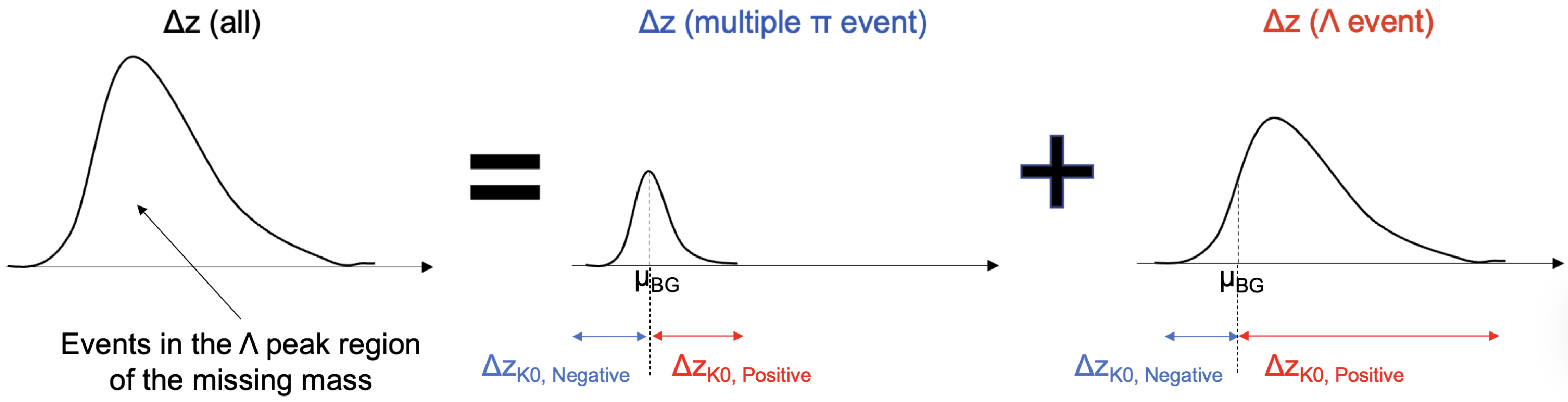}
    \caption{Schematic view of $\dzkz$ of events in the range of $1.07<\mm<1.16$ GeV/$c^{2}$. Two components are contained in the $\dzkz$ distribution. The first component is the background contribution with a symmetric structure. The second one is the $\Kz$ contribution having an exponential tail.}
    \label{fig-sch_dz}
\end{figure}

\begin{figure}[!h]
    \centering
    \includegraphics[width=0.75\columnwidth]{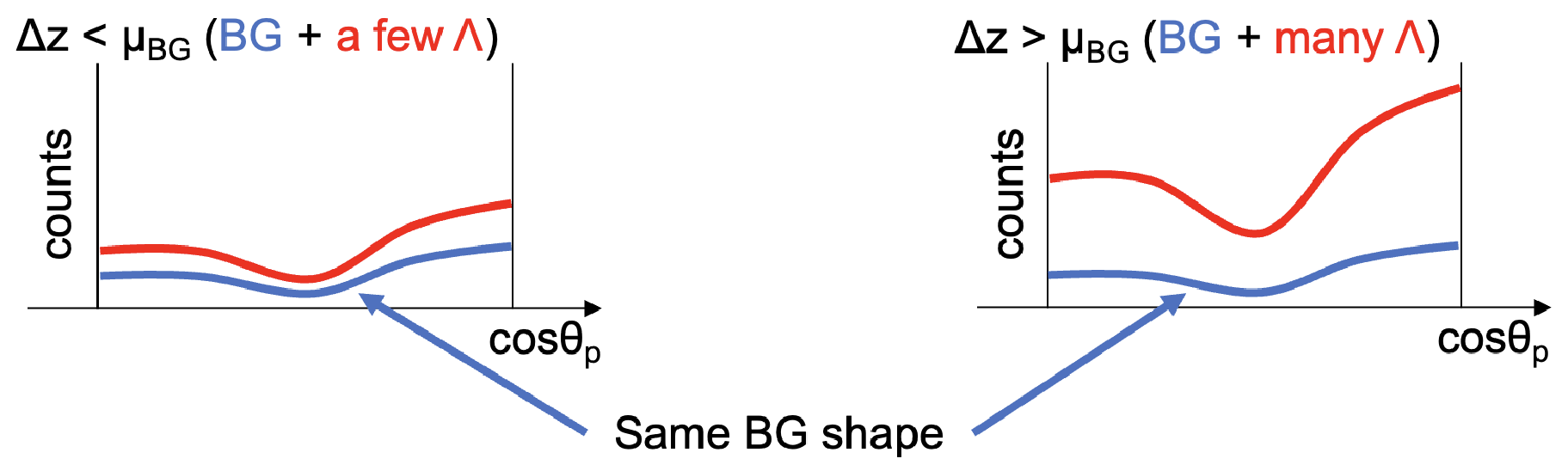}
    \caption{Schematic view of the $\costp(\dz<\mubg)$ and $\costp(\dz>\mubg)$ side. The $\costp$ \fix{distribution attributed to the background events} has the same structure on both sides. In contrast, the contribution of $\Lam$ production is more included in \fix{$\costp(\dz>\mubg)$}.}
    \label{fig-sch_costp}
\end{figure}

The events selected in \fix{Fig.} \ref{fig-mm_aftercut} still include \fix{both $\Lam$ productions (peak in Fig. \ref{fig-mm_aftercut}) and} the $\pP\pM$ pair from background events (blue shade in \fix{Fig.} \ref{fig-mm_aftercut}). \fix{The $\dzkz$ distribution calculated for these events includes both the exponential distribution representing true $\Kz$ flight distance and a symmetric distribution around zero due to the background}, as schematically shown in \fix{Fig.} \ref{fig-sch_dz}. \fix{The important point is that the background distribution has the symmetric structure for the peak position (hereafter referred to as $\mubg$). As shown in Fig. \ref{fig-sch_costp}, the $\costp$ distribution for $\dz<\mubg$ ($\costp(\dz<\mubg)$) and that for $\dz>\mubg$ ($\costp(\dz>\mubg)$) contain the same amount of background contribution, whereas $\costp(\dz>\mubg)$ contains more $\Lam$ events. Therefore, by subtracting $\costp(\dz<\mubg)$ from $\costp(\dz>\mubg)$, the pure angular distribution \fixxx{resulting} from $\Lam$ can be obtained. This subtraction also removes parts of the $\Lam$ event. However, it does not distort the angular distribution.}

\begin{figure}[h]
  \centering
  \includegraphics[width=\columnwidth]{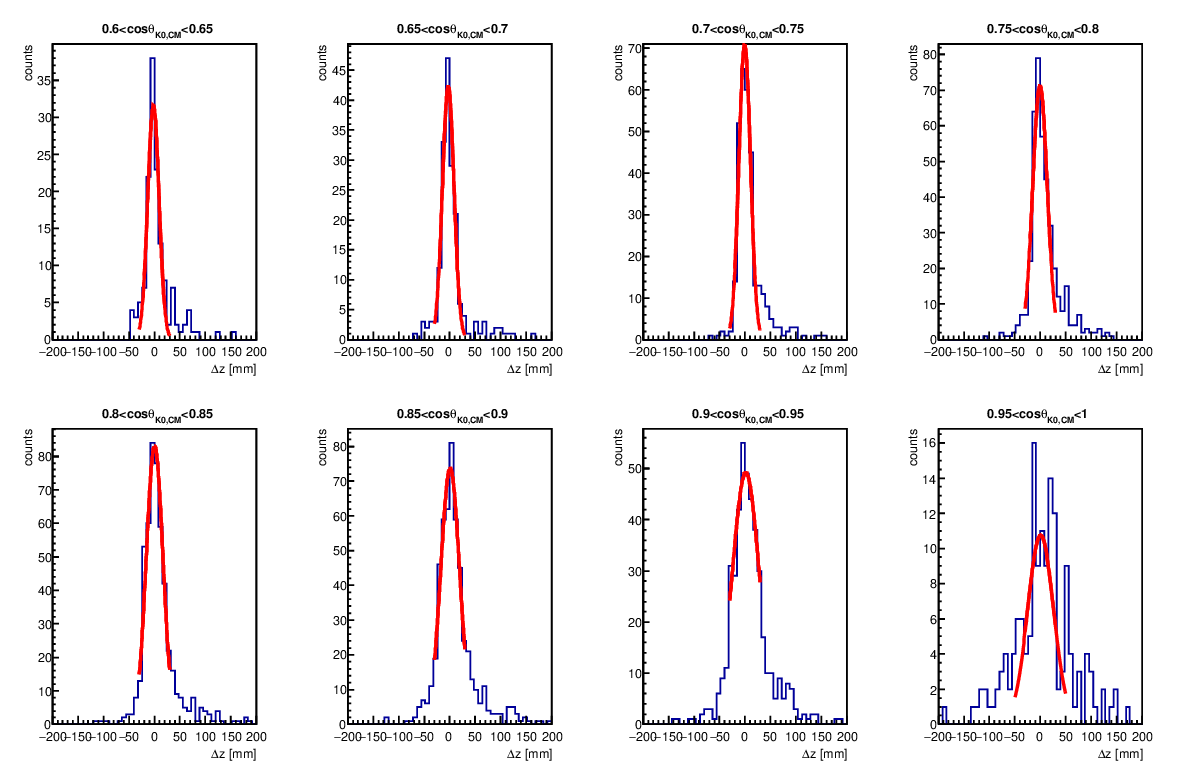}
  \caption{$\dzkz$ \fix{distribution for} events in the range of $1.0<\mm<1.07$ GeV/$c^{2}$. The red solid lines represent the fitting function (Gauss) to \fix{determine} the mean, $\mubg$.}
  \label{fig-dz_lowSB_fit}
\end{figure}

To show this analysis is valid, we must confirm that the $\dz$ distribution is symmetric for background events. In addition, the peak position should be determined from the analysis. To achieve this, we obtained the $\dz$ distribution for the events in the range of $1.0<\mm<1.07$ GeV/$c^{2}$ as shown in Fig. \ref{fig-dz_lowSB_fit}, where the $\costkz$ range was 0.6-1.0 with an angular bin width of $d\costkz = 0.05$. 
\wip{Since the background events are dominant in this missing mass range, this $\dz$ distribution is symmetric as expected.} \wip{These histograms were fitted with a Gaussian to obtain $\mubg$. The value of $\mubg$ does not change significantly when the fit range is varied.} 

\begin{figure}[h]
  \centering
  \includegraphics[width=\columnwidth]{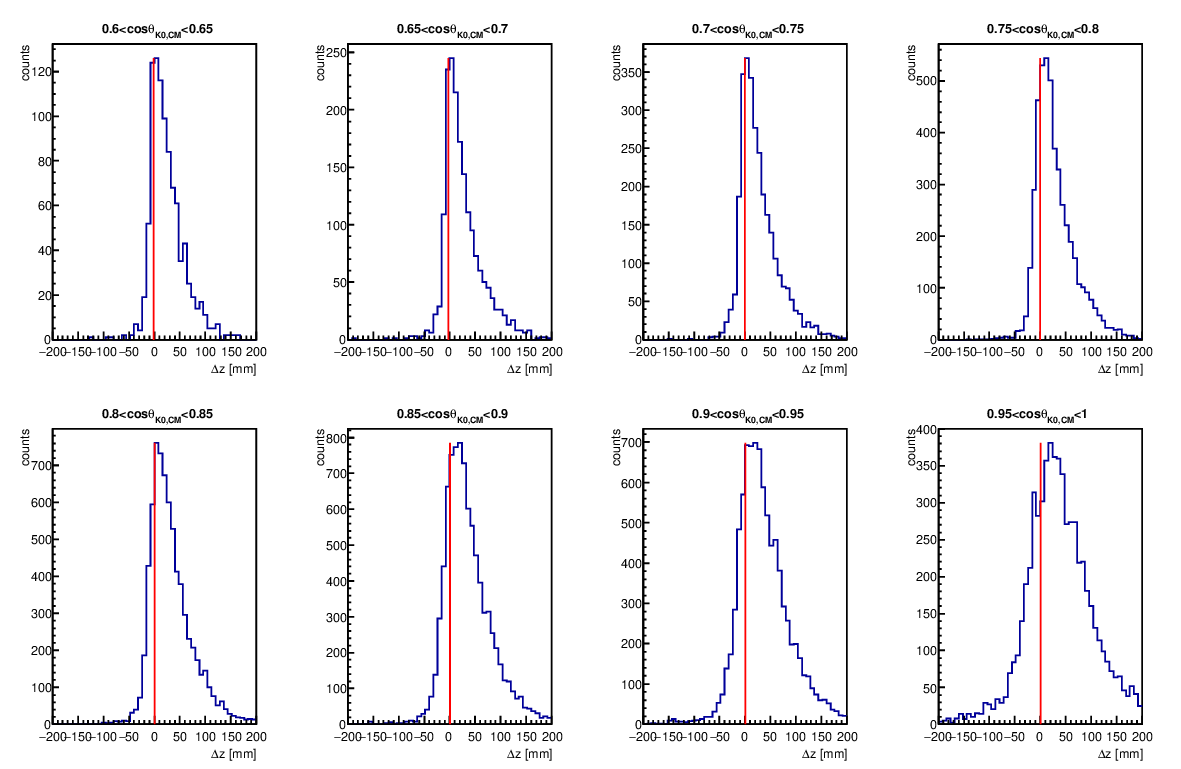}
  \caption{$\dzkz$ \fix{distribution for events in the range of $1.07<\mm<1.16$ GeV/$c^{2}$}. The red solid lines represent $\mubg$ obtained in Fig. \ref{fig-dz_lowSB_fit}.}
  \label{fig-dz_lam}
\end{figure}

Fig. \ref{fig-dz_lam} shows the $\dzkz$ distribution for events of \wip{$1.07<M_{\PiKX}<1.16$} GeV/$c^{2}$ in Fig. \ref{fig-mm_aftercut} for the $\costkz$ region of 0.6-1.0 with an angular bin width of $d\costkz=0.05$. Unlike the background events, the $\Kz$ flight length for these events \fixxx{exhibits} an asymmetric distribution attributed to true $\Kz$ events. The red solid lines represent $\mubg$. We should remember that the background contribution exists symmetrically around $\mubg$.

\begin{figure}[!h]
    \centering
    \includegraphics[width=\columnwidth]{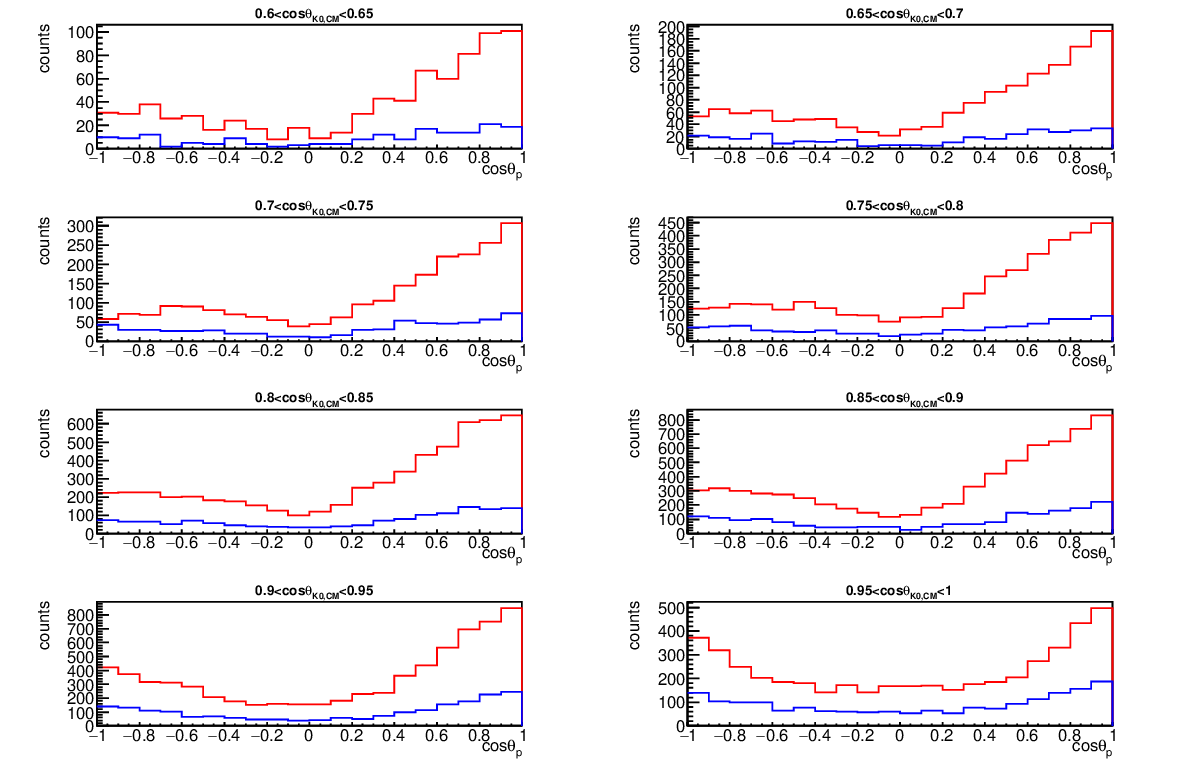}
    \caption{\fix{$\costp(\dz>\mubg)$} (red) and \fix{$\costp(\dz<\mubg)$} (blue) distributions. The events used here are those in the range of $1.07<\mm<1.16$ GeV/$c^{2}$.}
    \label{fig-costp_ab}
\end{figure}

\begin{figure}[h]
  \centering
  \includegraphics[width=\columnwidth]{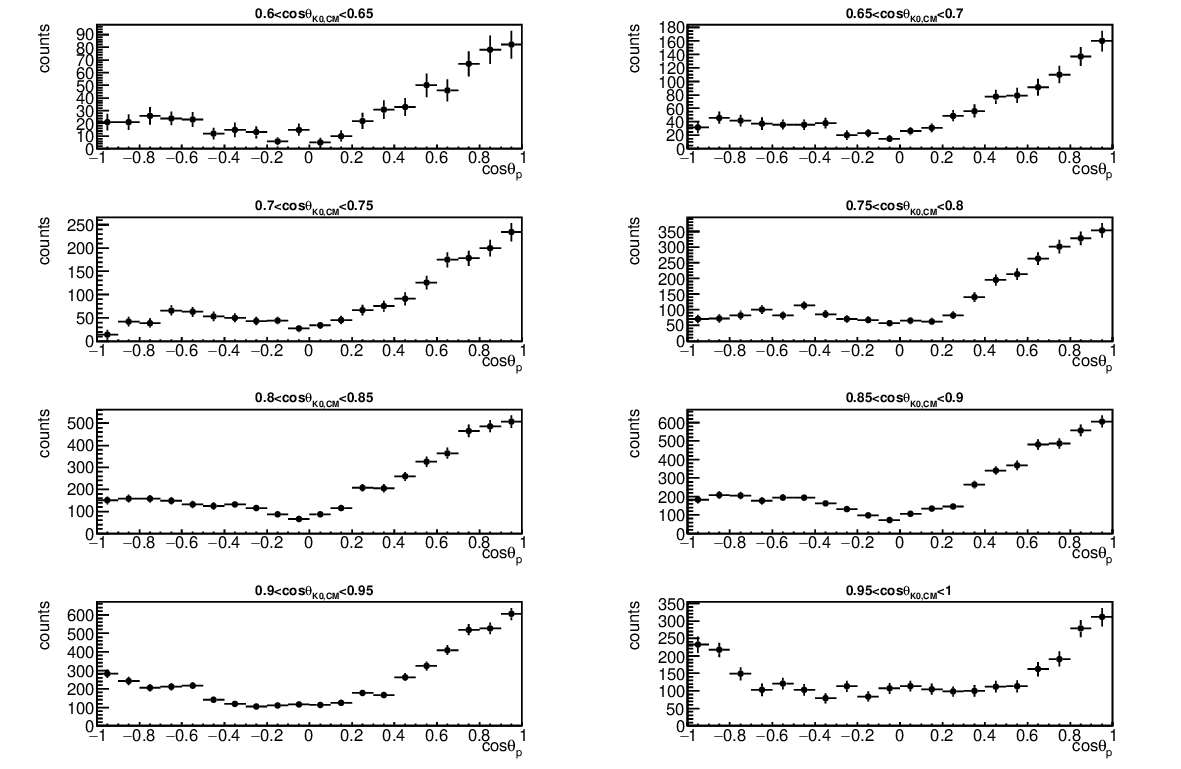}
  \caption{Extracted $\costp$ \fix{for the pure $\Lam$ events}.}
  \label{fig-costp_ext}
\end{figure}

\fix{Now, we are ready to extract the $\costp$ distribution for pure $\Lam$ events by subtracting $\costp(\dz<\mubg)$ from $\costp(\dz>\mubg)$, which are shown by blue and red histograms in Fig. \ref{fig-costp_ab}, respectively. Extracted $\costp$ distribution for pure $\Lam$ events is shown in Fig. \ref{fig-costp_ext}.}

%% file: analysis4.tex
\section{\bf Analysis \rom{4}: Correction of the Extracted $\costp$ Distribution}
\label{sec-correction}
\label{sec-ana4}

\begin{figure}[h]
  \centering
  \includegraphics[width=\columnwidth]{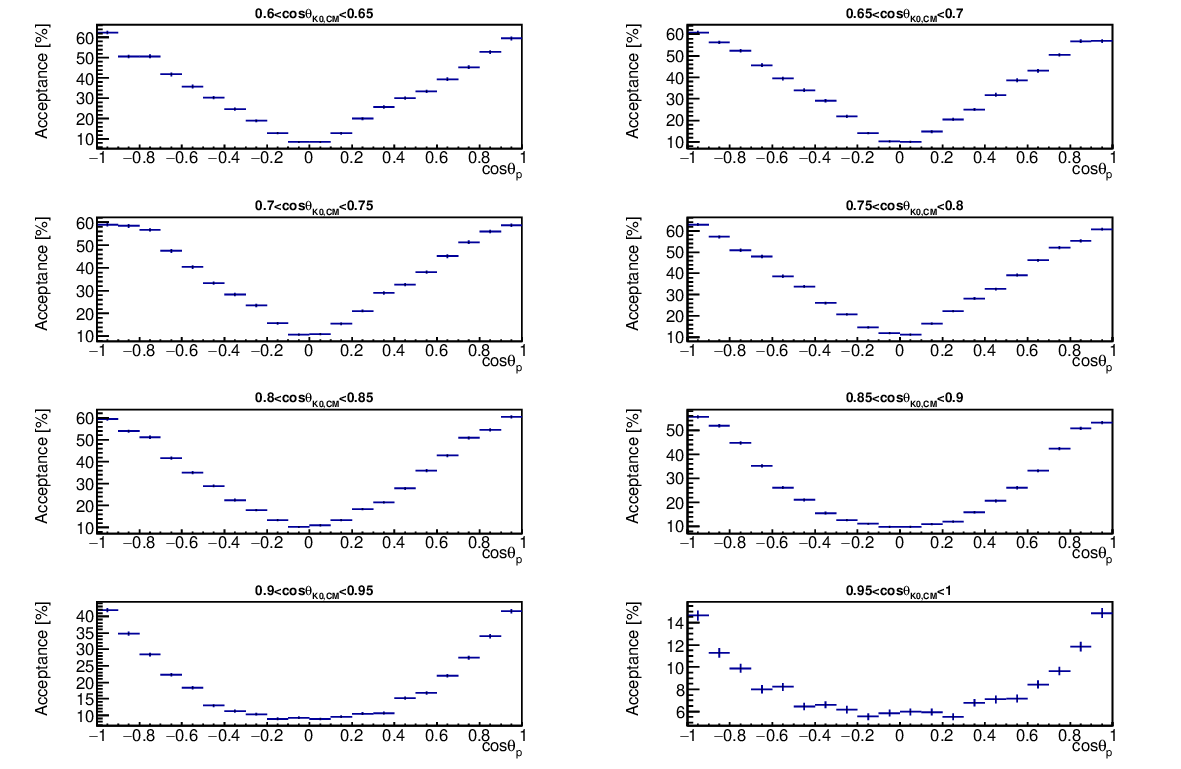}
  \caption{Estimated CTACH acceptance \fix{as a function of} $\costp$ for $\Lam$ production followed by the $\Ldecay$ decay.}
  \label{fig-accPl}
\end{figure}

The obtained $\costp$ (Fig. \ref{fig-costp_ext}) is influenced by the acceptance of CATCH and has a dip \wip{around $\costp=0$.} To correct these $\costp$ before deriving $P_{\Lam}$, the CATCH acceptance for the $\Ldecay$ decay was estimated using a Geant4-based Monte Carlo simulation. The $\PiKL$ reaction, followed by the $\Ldecay$ decay with a realistic branching ratio of 63.9\%, was generated. Here, CATCH detection efficiency \cite{Miwa-SMp}, which includes the detector acceptance, the tracking efficiency of CFT, and the energy measurement efficiency of BGO, was considered. The number of extracted events in the $i$-th $\costp$ bin was corrected by the CATCH acceptance as
\begin{align}
  N_i &= \frac{N_{0,\ i}}{\accPLi}, \\
  \sigma(N_i) &= N_i \sqrt{\left( \frac{\sigma(N_{0,\ i})}{N_{0,\ i}} \right)^2 +\left( \frac{(\erraccPLi}{\accPLi} \right)^2 },  \label{eq-errNcor} \\
\end{align}
where $N_i$ is the corrected yield, $\sigma(N_i)$ is the statistical error included in $N_i$. Fig. \ref{fig-accPl} shows the estimated CATCH acceptance. 

%% file: analysis5.tex
\section{\bf Analysis \rom{5}: Derivation of $\Lam$ Polarization}
\label{sec-chisq}
\label{sec-ana5}

The $\costp$s corrected \fixxx{using} the CATCH acceptance are shown in Fig. \ref{fig-costp_cor}. The corrected $\costp$s \fixxx{exhibit} the expected distribution with almost linear angular dependence for the $\Kz$ angular region of $0.6<\costkz<0.8$ and $0.95<\costkz<1.0$. The $P_{\Lam}$ can be derived by fitting these $\costp$s with a linear function corresponding to Equation (\ref{eq-costp}). The linear function was defined as
\begin{equation}
  f(x) = \frac{p_0}{2} (1+p_{1}x),
  \label{eq-fitfunc}
\end{equation}
where the fitting parameter $p_0$ corresponds to the total counts of the corrected $\costp$ (i.e., $N_0$ in Equation (\ref{eq-costp})) and $p_1$ corresponds to $\alpha P_{\Lambda}$ in Equation (\ref{eq-costp}). 

In contrast, analysis for the $\Kz$ angular region of $0.8<\costkz<0.95$ is not as straightforward, because the unexpected enhancement can be observed around central regions from $\costp=0$ to $\costp=\pm0.4$. In these regions, the protons are emitted in the beam direction, and protons are detected around the detector acceptance boundary of CATCH. The understanding of the CATCH efficiency, including this acceptance boundary, may not be perfect. \fixxx{Consequently}, this $\costp$ region is expected to be overcorrected and the $\costp$ distribution distorted.

\subsection{\wip{Data selection using $\chisq$ test}}
\begin{figure}[h]
  \centering
  \includegraphics[width=\columnwidth]{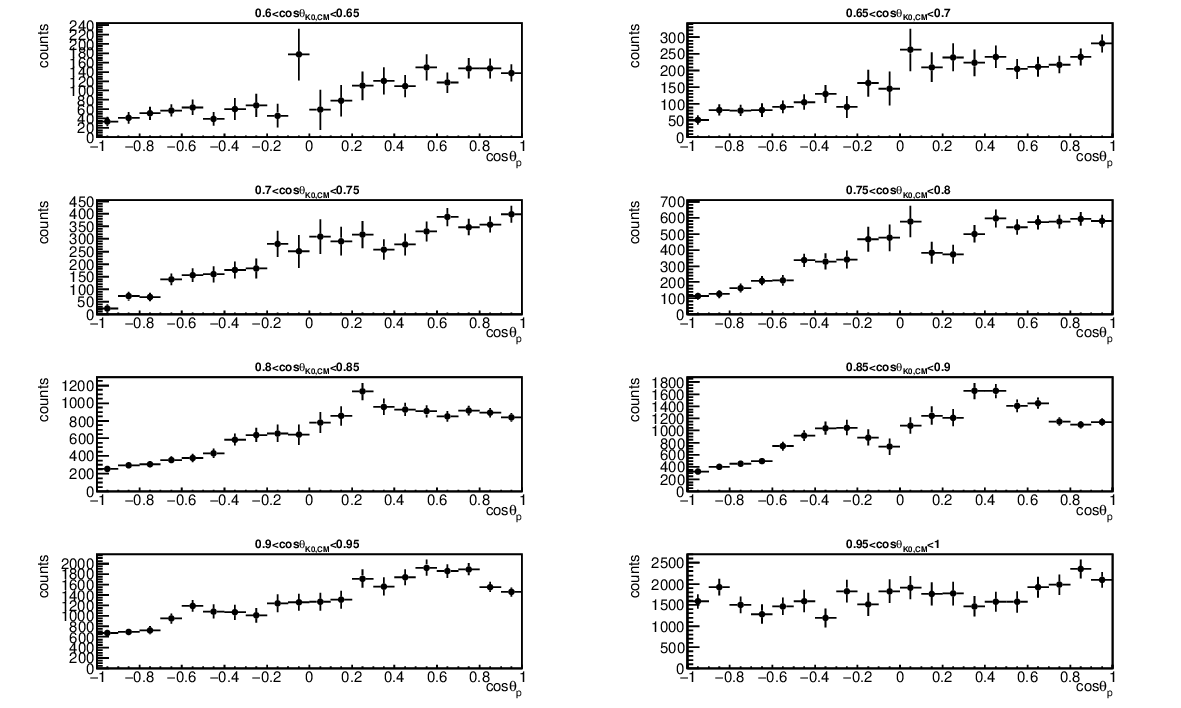}
  \caption{Corrected $\costp$ \fix{for the pure $\Lam$ events}.}
  \label{fig-costp_cor}
\end{figure}

\wip{In Fig. \ref{fig-costp_cor}, the uncertainty of CATCH acceptance correction made the $\costp$ distribution deviate from the expectation, especially in the $0.8<\costkz<0.95$ range. To select datapoints optimal for fitting to, a $\chisq$ test was performed in each $\costkz$ range with an interval of $d\costkz=0.05$.}

\begin{figure}[!h]
  \centering
  \includegraphics[width=15cm]{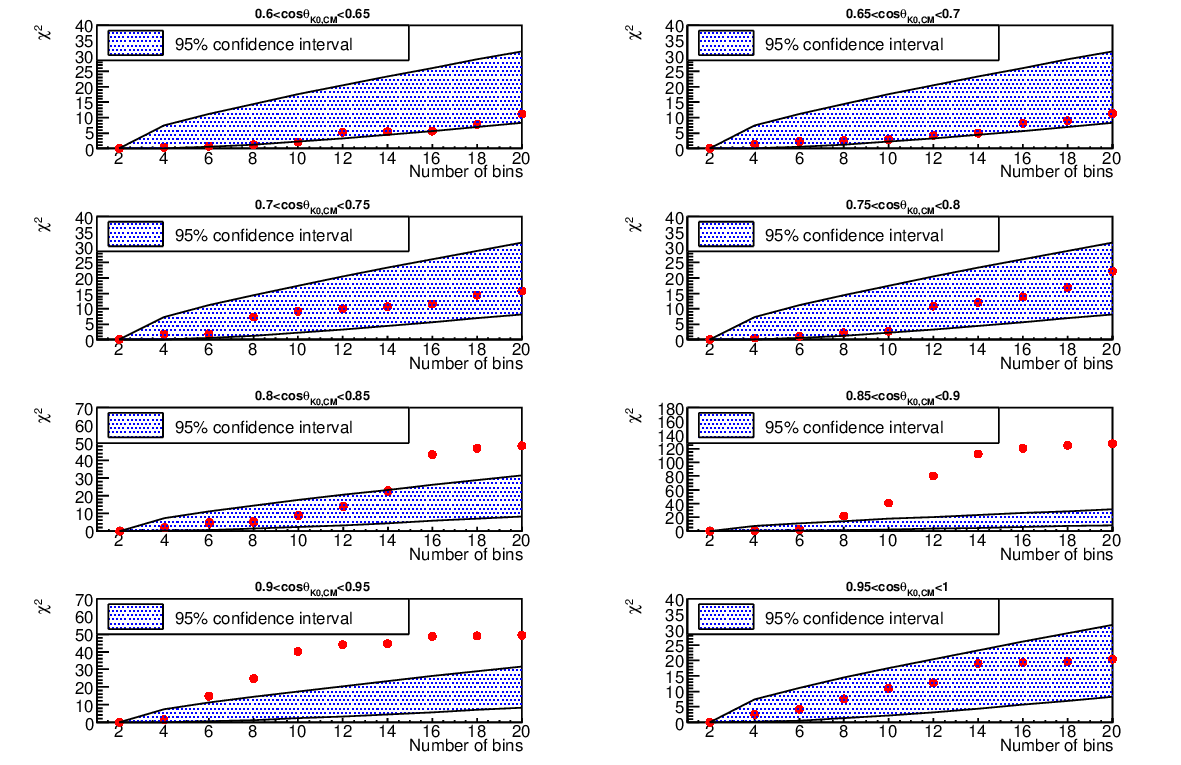}
  \caption{$\chisq$ of the fitting in each degree of freedom (red square point). The blue-shaded band represents the 95\% confidence interval.}
  \label{fig-chi2test}
\end{figure}

\begin{figure}[h]
  \centering
  \includegraphics[width=\columnwidth]{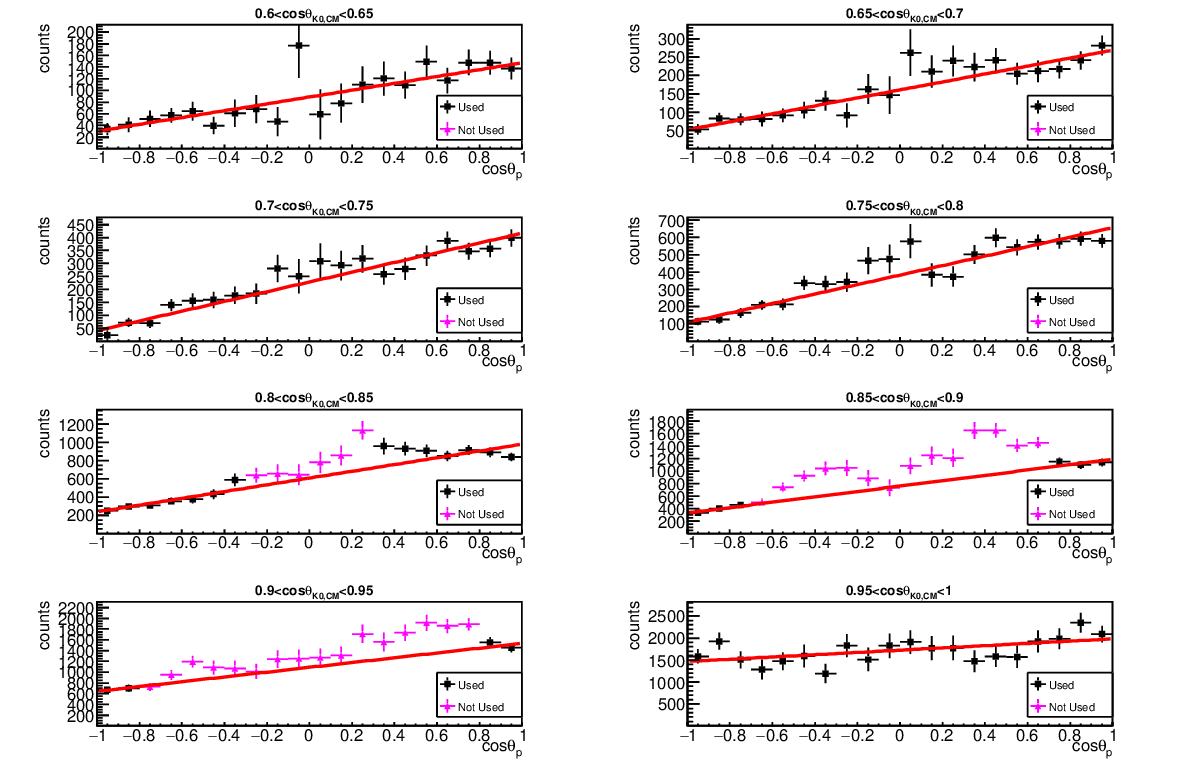}
  \caption{Corrected $\costp$ with the normalized angular region, showing the accepted data points (black), rejected ones (magenta), and the fitting function (red solid line).}
  \label{fig-costp_fit_merge}
\end{figure}

\wip{In general, if $N$ independent variables $x_i$ are each normally distributed with mean $\mu_i$ and variance $\sigma_i^2$, then the quantity known as $\chi^2$ is defined by
\begin{equation}
  \chi^2 \equiv \frac{(x_1 - \mu_1)^2}{\sigma_1^2} + \frac{(x_2 - \mu_2)^2}{\sigma_2^2} + \dots + \frac{(x_N - \mu_N)^2}{\sigma_N^2} = \sum_{i=1}^N \frac{(x_i - \mu_i)^2}{\sigma_i^2}.
\end{equation}
In our case, we re-defined that $x_i$ is the corrected $\costp$ counts ($N_i$), $\mu_i$ is the fitting result, and $\sigma_i$ is the error of corrected $\costp$ yield ($\sigma(N_i)$). The largest degree of freedom where the $\chisq$ falls within the 95\% confidence interval was selected. Here, we changed the degree of freedom by reducing the data points by two from either side of the center. Since the number of fitting parameters is two as in Equation (\ref{eq-fitfunc}), the optimized degree of freedom ($N_{dof}$) can be defined by the number of bins for fitting ($N_{bin}$) as 
\begin{equation}
  N_{dof} = N_{bin} - 2.
\end{equation}
The $\chisq$ with each $N_{bin}$ is shown in Figure \ref{fig-chi2test}. The red square point represents the obtained $\chisq$, and the blue-shaded band represents the 95\% confidence interval. The numerical values of the optimized $N_{bin}$ in each $\costkz$ range are summarized in Table \ref{tab-selebin}. Since $P_{\Lam}$ is synonymous with the slope of the $\costp$ distribution, $P_{\Lam}$ can be determined more accurately with this optimization. Fig. \ref{fig-costp_fit_merge} shows the accepted data points (black) and rejected ones (magenta), with the fitting functions (red solid line).}

\wip{Consequently, $P_{\Lam}$ and its statistical error ($\delta_{stat,\ j}$) in the $j$-th $\costkz$ bin can be expressed as
\begin{align}
  P_{\Lam,\ j} &= \frac{p_{1,\ j}}{\alpha}, \label{eq-PL} \\
  \delta_{stat,\ j} &= P_{\Lam,\ j} \sqrt{ \left( \frac{\delta(p_{1,\ j})}{p_{1,\ j}} \right)^2 + \left( \frac{\delta(\alpha)}{\alpha} \right)^2},
\end{align}
where $\delta(p_{1,\ j})$ is the fitting error of $p_1$ and $\delta(\alpha)=0.009$ \cite{Alpha}. }

\begin{table}[!h] 
  \begin{center}
  \caption{The number of bins in each $\costkz$ range selected by the $\chisq$ test (see Figure \ref{fig-chi2test}).}
  \centering
  \begin{threeparttable}
    \begin{tabular}{cc}
    $\costkz$ & Selected $N_{bin}$ \\
    \midrule\midrule
    $0.6-0.65$ & 20 \\
    \midrule
    $0.65-0.7$ & 20 \\
    \midrule
    $0.7-0.75$ & 20 \\
    \midrule
    $0.75-0.8$ & 20 \\
    \midrule
    $0.8-0.85$ & 14 \\
    \midrule
    $0.85-0.90$ & 6 \\
    \midrule
    $0.90-0.95$ & 4 \\
    \midrule
    $0.95-1.0$ & 20 \\
    \end{tabular}
  \end{threeparttable}
  \label{tab-selebin}
  \end{center}
\end{table}

%% file: systematic.tex
\section{\bf Systematic Uncertainty}
\label{sec-sysun}

The systematic uncertainties of $P_{\Lam}$ are primarily caused by the following factors:
\begin{itemize}
  \item {\bf The cut range dependence for the $\Ldecay$ decay selection}\\
  We identified the $\Ldecay$ decays by investigating whether the $\Lam$ momenta obtained by production and decay kinematics are identical. Here, we used the index \say{$\Delta p$}, as described in Sec. \ref{sec-AnaLdecay}. The systematic uncertainty from the \fix{selection} of the $\Ldecay$ decay was evaluated by changing the cut value for $\Delta p$ from (-0.06, 0.04) to (-0.05, 0.03). \wip{As mentioned in Sec. \ref{sec-ana2}, since the central value of the $\Delta p$ distribution is off due to the effect of insufficient energy calibration of BGO and CFT, an interval of 0.1 GeV/$c$ was chosen to apply the $\Delta p$ resolution symmetrically from the actual center.}\\  
  \item {\bf The uncertainty of background cancelation in $\costp$ distribution}\\
  We subtracted the background in $\costp$, \fixxx{primarily} derived from multiple $\pi$ production, by referring to the $\Kz$ \fix{flight} length in the beam axis ($\dzkz$). Here, \fix{we estimated the mean value ($\mubg$) of the $\dzkz$ distribution for the background events in the missing mass spectrum}. \fixxx{Thereafter}, we subtracted the \fix{$\costp(\dz<\mubg)$} events from the $\costp(\dz>\mubg)$ events, as shown in \fix{Fig.} \ref{fig-sch_costp}. This background cancellation performance depends on the $\mubg$ value. The systematic uncertainty \fix{due to the imperfection of} background \fix{cancelation} was evaluated by changing the value of \wip{$\mubg$ to $\mu_{BG}\pm\sigma$, where $\sigma$ is the standard deviation of the fitting function in Fig. \ref{fig-dz_lowSB_fit}.} \wip{Therefore, it could be said that the systematic error from varying the} \wip{fitting range of $\dz$ distribution is a conservative estimate.}\\
  \item {\bf The data sample dependence used at the $\costp$ fitting}\\
  \fix{We selected the data sample in the fitting of $\costp$ distribution to remove the overcorrected data point.} The systematic uncertainty \fix{depending on the data selection} was evaluated by fitting $\costp$ with all data points and remeasured $P_{\Lam}$. \\
\end{itemize}

We estimated the relative error $\delta_{rel}$ caused by each factor above using the absolute error $\delta_{abs}$ of each as
\begin{align}
\delta_{rel} &= \frac{P_{\Lam,\ nominal} - P_{\Lam,\ variation}}{P_{\Lam,\ nominal}}, \\
		  &= \frac{\delta_{abs}}{P_{\Lam,\ nominal}},
\end{align}
where $P_{\Lam,\ nominal}$ is the final result, $P_{\Lam,\ variation}$ is the polarization value derived by changing each value of systematic uncertainty factors. Table \ref{tab-syserr} shows the estimated relative errors in each $\costkz$ bin. Here, $\delta_{rel,\ purity}$, $\delta_{rel,\ \mubg+\sigma}$, $\delta_{rel,\ \mubg-\sigma}$, $\delta_{rel,\ fitting}$ represent the relative systematic errors contributing to the purity of the $\Ldecay$ decay events, the uncertainty in the background estimatation of the $\costp$ distribution, and the uncertainty in the fitting of the $\costp$ distribution, respectively. The reason why $\delta_{rel,\ fitting}$s in the ranges of $0.6<\costkz<0.65$, $0.65<\costkz<0.7$, $0.7<\costkz<0.75$, $0.75<\costkz<0.8$, and $0.95<\costkz<1.0$ are zero is that we used all data points for the fitting in these ranges according to the $\chisq$ test described in Sec. \ref{sec-chisq}.

The total systematic error in the $j$-th $\costkz$ bin $\delta_{sys,\ j}$ was calculated as
\begin{gather}
  \delta_{sys,\ j} = \sqrt{\delta_{abs,\ purity,\ j}^{2} + \delta_{abs,\ \mubg+\sigma,\ j}^{2} + \delta_{abs,\ \mubg-\sigma,\ j}^{2} + \delta_{abs,\ fitting,\ j}^{2}} \\
  \nonumber \\
  \delta_{abs,\ purity,\ j} = |P_{\Lam,\ nominal,\ j} - P_{\Lam,\ purity,\ j}| \\
  \delta_{abs,\ \mubg+\sigma,\ j} = |P_{\Lam,\ nominal,\ j} - P_{\Lam,\ \mubg+\sigma,\ j}| \\
  \delta_{abs,\ \mubg-\sigma,\ j} = |P_{\Lam,\ nominal,\ j} - P_{\Lam,\ \mubg-\sigma,\ j}| \\
  \delta_{abs,\ fitting,\ j} = |P_{\Lam,\ nominal,\ j} - P_{\Lam,\ fitting,\ j}| 
\end{gather}
where $P_{\Lam,\ purity,\ j}$ is the polarization value obtained by changing the selection of $\Ldecay$ decay (i.e., $\Delta p$ distribution) in Fig. \ref{fig-dp}, $P_{\Lam,\ \mubg\pm\sigma,\ j}$ is the one obtained by changing the background contamination (i.e., fitting frunctuation of $\dz$ distributions in Fig. \ref{fig-dz_lowSB_fit}), and $P_{\Lam,\ fitting,\ j}$ is the one obtained by changing the degree of freedom for the $\costp$ fitting. The numerical value of the total systematic error is displayed in Table \ref{tab-Pl}. 

\begin{table}[!h] 
  \begin{center}
  \caption{Estimated relative errors in each $\costkz$ bin.}
    \begin{tabular}{ccc}
    $\costkz$ & Candidate & Relative Error (\%) \\ \hline\hline
    0.6-0.65 & 
    \begin{tabular}{c}
    $\delta_{rel,\ purity}$ \\ $\delta_{rel,\ \mubg+\sigma}$ \\ $\delta_{rel,\ \mubg-\sigma}$ \\ $\delta_{rel,\ fitting}$
    \end{tabular} &
    \begin{tabular}{c}
    0.37 \\ 2.28 \\ 2.55 \\ 0.00
    \end{tabular} \\ \hline

    0.65-0.7 & 
    \begin{tabular}{c}
    $\delta_{rel,\ purity}$ \\ $\delta_{rel,\ \mubg+\sigma}$ \\ $\delta_{rel,\ \mubg-\sigma}$ \\ $\delta_{rel,\ fitting}$
    \end{tabular} &
    \begin{tabular}{c}
    0.82 \\ 3.46 \\ 1.22 \\ 0.00
    \end{tabular} \\ \hline
    
    0.7-0.75 & 
    \begin{tabular}{c}
    $\delta_{rel,\ purity}$ \\ $\delta_{rel,\ \mubg+\sigma}$ \\ $\delta_{rel,\ \mubg-\sigma}$ \\ $\delta_{rel,\ fitting}$
    \end{tabular} &
    \begin{tabular}{c}
    1.22 \\ 0.01 \\ 0.35 \\ 0.00
    \end{tabular} \\ \hline

    0.75-0.8 & 
    \begin{tabular}{c}
    $\delta_{rel,\ purity}$ \\ $\delta_{rel,\ \mubg+\sigma}$ \\ $\delta_{rel,\ \mubg-\sigma}$ \\ $\delta_{rel,\ fitting}$
    \end{tabular} &
    \begin{tabular}{c}
    3.38 \\ 2.32 \\ 1.41 \\ 0.00
    \end{tabular} \\ \hline

    0.8-0.85 & 
    \begin{tabular}{c}
    $\delta_{rel,\ purity}$ \\ $\delta_{rel,\ \mubg+\sigma}$ \\ $\delta_{rel,\ \mubg-\sigma}$ \\ $\delta_{rel,\ fitting}$
    \end{tabular} &
    \begin{tabular}{c}
    1.38 \\ 1.09 \\ 1.73 \\ 0.19
    \end{tabular} \\ \hline    

    0.85-0.9 & 
    \begin{tabular}{c}
    $\delta_{rel,\ purity}$ \\ $\delta_{rel,\ \mubg+\sigma}$ \\ $\delta_{rel,\ \mubg-\sigma}$ \\ $\delta_{rel,\ fitting}$
    \end{tabular} &
    \begin{tabular}{c}
    1.48 \\ 1.33 \\ 1.86 \\ 2.43
    \end{tabular} \\ \hline

    0.9-0.95 & 
    \begin{tabular}{c}
    $\delta_{rel,\ purity}$ \\ $\delta_{rel,\ \mubg+\sigma}$ \\ $\delta_{rel,\ \mubg-\sigma}$ \\ $\delta_{rel,\ fitting}$
    \end{tabular} &
    \begin{tabular}{c}
    3.91 \\ 0.74 \\ 0.09 \\ 9.11
    \end{tabular} \\ \hline

    0.95-1.0 & 
    \begin{tabular}{c}
    $\delta_{rel,\ purity}$ \\ $\delta_{rel,\ \mubg+\sigma}$ \\ $\delta_{rel,\ \mubg-\sigma}$ \\ $\delta_{rel,\ fitting}$
    \end{tabular} &
    \begin{tabular}{c}
    1.83 \\ 0.95 \\ 5.88 \\ 0.00
    \end{tabular} \\ \hline
     
    \label{tab-syserr} 
    \end{tabular}
  \end{center}
\end{table}

%% file: results.tex
\section{\bf Results}
\label{sec-result}

\begin{figure}[h]
  \centering
  \includegraphics[width=0.7\columnwidth]{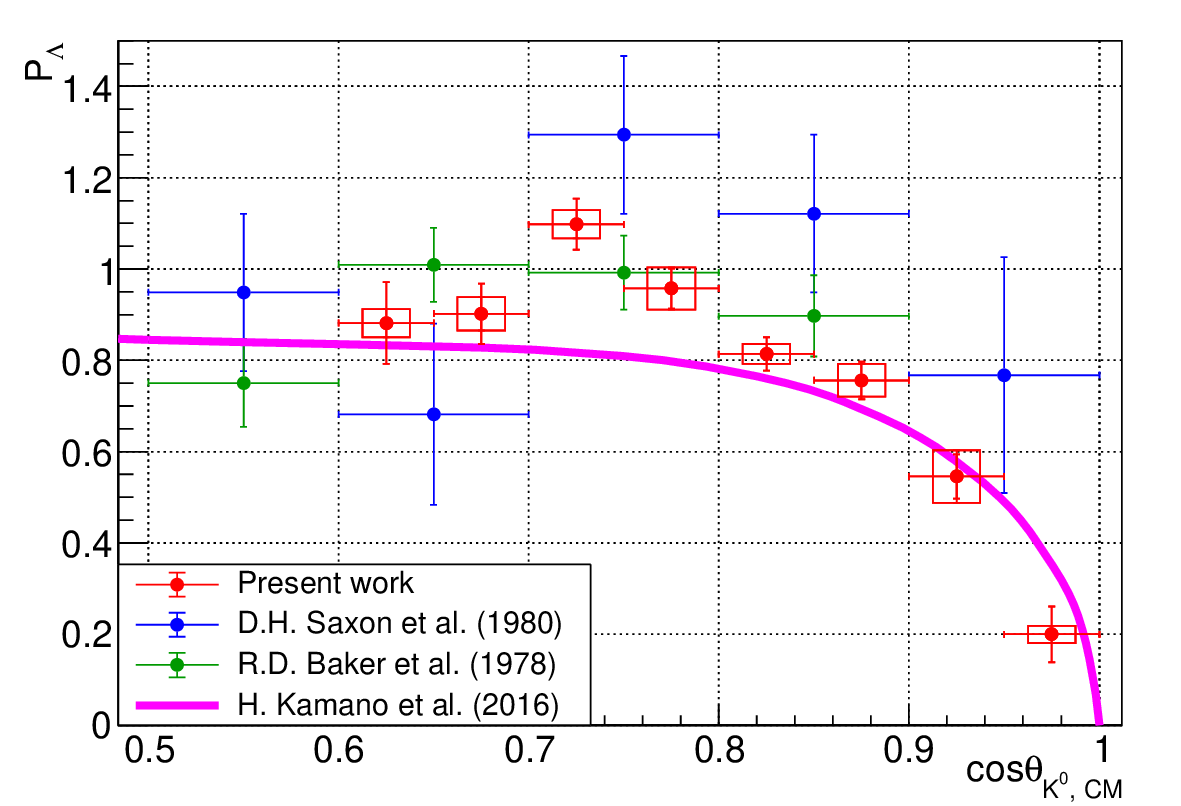}
  \caption{$\Lambda$ polarizations in the $\PiKL$ reaction, in the present work (\fix{the} red-filled circles). \fix{The error bars and boxes show the statistical and systematic errors. Past data (}Ref. \cite{Baker-1978} (green-filled circles), and Ref. \cite{Saxon-1980} (blue-filled circles)\fix{) are shown together}. The magenta solid line represents the result from \fix{a} PWA calculation \cite{Kamano-2016}.}
  \label{fig-Pl}
\end{figure}

Finally, the $\Lambda$ polarization ($P_{\Lambda}$) in the $\PiKL$ reaction was \fix{derived} using Equation (\ref{eq-PL}). \fix{Fig.} \ref{fig-Pl} shows the results of the present work (\fix{the} red-filled circles) with the statistical (bar) and systematic (box) errors. Past data are overlaid here \fixxx{using} \fix{the} green-filled circles \cite{Baker-1978} and blue-filled circles \cite{Saxon-1980}. Both \fix{the} statistical and systematic errors \fix{are significantly} improved compared \fixxx{with} the past measurements. The numerical values included in this measurement are summarized in Table \ref{tab-Pl}. The averaged $P_{\Lambda}$ value in the $\costkz$ range of 0.6-0.85 was $0.932\pm0.058\pm0.028$, which means \fixxx{that} $\Lambda$ produced by the $\PiKL$ reaction is highly polarized in this angular range. By selecting such a specific range, we can use the highly polarized $\Lam$ beam for the $\Lp$ spin observable measurement in J-PARC E86. 

\begin{table}[!h] 
  \begin{center}
  \caption{Numerical values included in the $P_{\Lam}$ measurement}
    \begin{tabular}{cccc}
    $\costkz$ & 
    \begin{tabular}{c}
    Baker $et\ al.$ \\ $P_{\Lambda}\pm\sigma$ 
    \end{tabular} &
    \begin{tabular}{c}
    Saxon $et\ al.$ \\ $P_{\Lambda}\pm\sigma$ 
    \end{tabular} &
    \begin{tabular}{c}
    Present work \\ $P_{\Lambda}\pm\sigma_{stat}\pm\sigma_{sys}$
    \end{tabular} 
    \\
    \hline\hline
    0.5 - 0.6 & $0.750\pm0.097$ & $0.949\pm0.173$ & \\
    \hline
    \begin{tabular}{c}
    0.6 - 0.65 \\ 0.65 - 0.7
    \end{tabular} & 
    $1.009\pm0.081$ & $0.682\pm0.198$ &
    \begin{tabular}{c}
    $0.885\pm0.087\pm0.030$ \\ $0.905\pm0.066\pm0.034$ 
    \end{tabular} \\
    \hline
    \begin{tabular}{c}
    0.7 - 0.75 \\ 0.75 - 0.8
    \end{tabular} &
    $0.992\pm0.081$ & $1.294\pm0.173$ &
    \begin{tabular}{c}
    $1.109\pm0.055\pm0.014$ \\ $0.951\pm0.044\pm0.041$ 
    \end{tabular} \\
    \hline
    \begin{tabular}{c}
    0.8 - 0.85 \\ 0.85 - 0.9
    \end{tabular} &
    $0.897\pm0.089$ & $1.121\pm0.173$ &
    \begin{tabular}{c}
    $0.812\pm0.036\pm0.020$ \\  $0.745\pm0.041\pm0.027$ 
    \end{tabular} \\
    \hline
    \begin{tabular}{c}
    0.9 - 0.95 \\ 0.95 - 1.0
    \end{tabular} &
     & $0.768\pm0.259$ &
     \begin{tabular}{c}
     $0.555\pm0.048\pm0.055$ \\ $0.189\pm0.061\pm0.012$
     \end{tabular} \\
    \label{tab-Pl}
    \end{tabular}
  \end{center}
\end{table}

%% file: discussion.tex
\section{\bf Discussion}
\label{sec-discussion}
\subsection{\bf Polarization mechanism in the $\PiKL$ reaction}
\wip{In the following, the discussion is based on the scattering theory described in Ref. \cite{Sasakawa-1991}. Considering the scattering of a spin 0 system and a spin 1/2 system, such as the scattering of $\pM$ mesons and protons, the spin matrix (scattering matrix) can be written as
\begin{equation}
M = f(\theta) + {\bm \sigma}\cdot{\bm n} g(\theta),
\label{eq-spinMatrix}
\end{equation}
where ${\bm \sigma}$ is the Pauli matrices, and ${\bm n}$ is the normal vector of the reaction plane defined by the momentum vectors (${\bm p}$ and ${\bm p'}$) in the initial and final states:
\begin{equation}
{\bm n} = \frac{{\bm p}\times{\bm p'}}{|{\bm p}\times{\bm p'}|}.
\end{equation}
In Equation (\ref{eq-spinMatrix}), the first term is called \say{non-spin flip term}, and the second is called \say{spin flip term}, respectively.}

\wip{If the initial state is not polarized, the final state polarization vector ${\bm P_{f}}$ can be written as
\begin{align}
{\bm P_{f}} 
&= \frac{{\rm Tr}(MM^{\dag}{\sigma})}{{\rm Tr}(MM^{\dag})} \\
&= \frac{(f^{*}(\theta)g(\theta) + f(\theta)g^{*}(\theta)) {\bm n}}{|f(\theta)|^{2} + |g(\theta)|^{2}}. \label{eq-Pf}
\end{align}
From the above, it can be said that the polarization of $\Lambda$ produced by the $\PiKL$ reaction can be described as an interference between a non-spin flip term and a spin flip term.} \wip{Equation (\ref{eq-Pf}) clearly shows that the polarization vanishes when either $f(\theta)$ or $g(\theta)$ is zero, as no interference is possible. Conversely, ${\bm P}_f$ is maximized when both amplitudes are non-zero, of comparable magnitude, and have a relative phase that results in a large real part of the product $f^*(\theta) g(\theta)$. For example, if $f(\theta)$ and $g(\theta)$ are real and equal in magnitude, the polarization reaches its theoretical maximum of unity.}

\wip{Therefore, the observed high polarization of the $\Lam$ hyperon in specific angular regions is not due to the dominance of the spin-flip amplitude alone, but rather to the constructive interference between the spin-flip and non-spin-flip components. This mechanism underscores the importance of partial-wave interference effects and highlights the sensitivity of hyperon polarization to the underlying resonance structure and angular momentum dynamics of the reaction.}

\subsection{\bf Comparison with theoretical models}
\fixxx{The} Bonn-Gatchina (BnGa) multichannel PWA group \cite{Bonn-PWA-2012} has \fixxx{indicated} that one possible reason \fixxx{for} the polarization of $\Lambda$ in the $\PiKL$ reaction \fixxx{is} the presence of the following nucleon resonances: $N(1710)\frac{1}{2}^{+}$ and $N(1900)\frac{3}{2}^{+}$. The theoretical calculation from Ref. \cite{Kamano-2016}, which investigated the nucleon resonances within a DCC model of $\pi N$ and $\gamma N$ reactions, shows a similar trend to our data. The DCC model they \fixxx{used} was parameterized in simultaneous fits to the data of $\pi N\to \eta N,\ K\Lam,\ K\Sigma$ and $\gamma N \to \pi N,\ \eta N,\ K\Lam,\ K\Sigma$ up to $W=2.1$ GeV and that of $\pi N\to \pi N$ up to $W=2.3$ GeV. While our $P_{\Lam}$ data are vital for fixing model parameters in these PWA studies, more spin observable data are necessary to discuss the specific reactions.

%% file: summary.tex
\section{\bf Summary}
\wip{In this paper, we precisely measured the $\Lam$ polarization ($P_{\Lam}$) in the $\PiKL$ reaction using the 1.33 GeV/$c$ $\pM$ beam to offer an essential input to the DCC models and to verify the feasibility of J-PARC E86. The observed average polarization in the region $0.60 < \costkz < 0.85$ was $0.932 \pm 0.058 (\text{stat}) \pm 0.028 (\text{syst})$. }

\wip{In our analysis, $\Kz$ was reconstructed from the decay $\pM$ and decay $\pP$, analyzed using CATCH and KURAMA spectrometer, respectively. Almost $3.56\times10^{4}$ $\Lambda$ particles were identified requiring the detection of $\pM$ and proton from the $\Ldecay$ decay.}

\wip{We analyzed the identified $\Ldecay$ decay events and derived the $P_{\Lam}$ using the angular distribution of decay proton ($\costp$) and how polarized the $\Lam$ is at each $\Kz$ scattering angle. From the present result, we confirmed that the spin observables measurement in J-PARC E86 is possible by selecting the specific angular range where $\Lam$ produced by the $\PiKL$ reaction has a high polarization. We also empirically confirmed that the observed high polarization is due to the constructive interference between the spin-flip and non-spin-flip components. We concluded that this mechanism underscores the importance of partial-wave interference effects and highlights the sensitivity of hyperon polarization to the underlying resonance structure and angular momentum dynamics of the reaction.} 
While our precise $P_{\Lam}$ data are vital for fixing model parameters in the DCC models, more spin observable data are necessary to discuss the specific reactions.  

\section*{\bf Acknowledgments}
We'd like to thank the staff of the J-PARC accelerator and Hadron Experimental Facility for their support in providing beam time. J-PARC E40 was supported by the Japan Society for the Promotion of Science KAKENHI Grants No. 25H01272, No. 24H00212, No. 23684011, No. 18H03693, No. 15H05442, No. 15H02079, and No. 15H00838. This experiment was also supported by Grants-in-Aid No. 24105003 and No. 18H05403 for Scientific Research from the Ministry of Education, Culture, Science and Technology (MEXT), Japan. This work was supported by the Japan Society for the Promotion of Science KAKENHI for DC1 Grants No. 21J20154 and No. 22KJ0166, and Graduate Program on Physics for the Universe (GP-PU) at Tohoku University. 

%% file: reference.tex
\vspace{10pt}

\let\doi\relax

%% file: end.tex
\end{document}